\documentclass[twocolumn,pre,english]{revtex4-1}
\usepackage[T1]{fontenc}
\usepackage{babel}
\usepackage{array}
\usepackage{float}
\usepackage{booktabs}
\usepackage{textcomp}
\usepackage{amsmath}
\usepackage{amsthm}
\usepackage{amssymb}
\usepackage{graphicx}
\usepackage{esint}
\usepackage{xcolor}
\usepackage{color}
\usepackage[unicode=true,pdfusetitle,
 bookmarks=true,bookmarksnumbered=false,bookmarksopen=false,
 breaklinks=false,pdfborder={0 0 1},backref=false,colorlinks=false]
 {hyperref}
 
\usepackage[sort&compress]{natbib}

\usepackage{silence}

\makeatletter

\providecommand{\tabularnewline}{\\}

\makeatother

\begin{document}
\title{Spectral Crossovers and Universality in Quantum Spin-chains Coupled to Random Fields}
\author{Debojyoti Kundu}
\email{debojyoti.kundu.physics@gmail.com}

\author{Santosh Kumar}
\email{Corresponding Author : skumar.physics@gmail.com}

\author{Subhra Sen Gupta}
\email{Corresponding Author : subhro.sengupta@gmail.com}

\affiliation{\textrm{\textit{Department of Physics, Shiv Nadar University, Gautam
Buddha Nagar, Uttar Pradesh 201314, India}}}

\begin{abstract}
We study the spectral properties of and spectral-crossovers between different
random matrix ensembles (Poissonian, GOE, GUE) in correlated spin-chain systems, in the presence of random magnetic fields, and the scalar
spin-chirality term, competing with the usual isotropic and time-reversal invariant Heisenberg term. We have investigated these crossovers in the context of the \emph{level-spacing
distribution} and the \emph{level-spacing ratio distribution}. We use random matrix theory (RMT) analytical results to fit the observed Poissonian-to-GOE and GOE-to-GUE  crossovers, and examine the relationship between the RMT crossover parameter $\lambda$ and scaled physical parameters of the spin-chain systems in terms of a scaling exponent. We find that the crossover behavior exhibits \emph{universality}, in the sense that it becomes independent of
lattice size in the large Hamiltonian matrix dimension limit.
\end{abstract}
\maketitle

\section{Introduction\label{sec:Introduction}}

The Hubbard model and some of its generalizations \cite{Hubbard_original,Fazekas's_book,Lieb_Hubbard_model,Mahan_many_particle_physics}
are prototypical models for interacting electrons moving  in a highly correlated manner, on a lattice. It is well known that under certain conditions like
large local (onsite) Coulomb repulsion and for an integer number of
electrons per lattice site, the $\mathit{charge}$ degrees of freedom
are frozen out (no net number and hence charge fluctuations at sites, to a first approximation) while the $\mathit{spin}$
and the $\mathit{orbital}$ degrees of freedom remain active via relative change
in intra-site spin-orbital occupancies. This results in effective
spin-orbital inter-site exchange interactions leading to Anderson
super-exchange \cite{Anderson_Super_exchange-1,Anderson_new_Superexchange,Anderson's_book}
and Kugel-Khomskii \cite{Kugel_Khomskii_1982,Khaliullin's_Kugel_Khomskii_model_paper}
like physics. For the single band Hubbard model, or for multi-band
models with very large Hund\textquoteright s coupling, orbital fluctuations are irrelevant and  this leads to a variety of spin-only Hamiltonians, the most basic of which is
the $\mathit{isotropic}$ $\mathit{Heisenberg}$ $\mathit{model}$.
While the parent fermionic models are difficult or impossible to solve
exactly, especially in higher dimensions, even the effective spin-Hamiltonians
embody very rich physics and often do not yield to exact solutions,
especially in the presence of spin-orbit coupling and magnetic field
induced anisotropies \cite{Moriya_DM_1,Moriya_DM_2,Dzyaloshinskii_DM,Spin_chirality_Wen_Wilczek,Spin_chirality_Rokhsar,Spin_chirality_Freericks,Diptiman's-paper}. Numerical solutions become very important in these circumstances. 

Another central paradigm of modern condensed matter physics is the presence
of disorder. While the problem of disorder (random site and bond energies mimicking impurities in real solids) in absence of interactions
was addressed a long time ago by Anderson and co-workers \cite{Anderson's_impurity,Scaling_theory_impurity_Abrahams_Anderson-1,Scaling_theory_impurity_Abrahams_Anderson-2,Fifty_years_of_Anderson_localization},
the generic problem of disorder in the presence of electron correlations
is a very difficult one and only limited success has been achieved
\cite{Efros_disordered_system_1975,Romer_Punnoose,Punnoose_Finkel'stein-1,Punnoose_Finkel'stein-2}.
In this regime one has to depend even more on numerical methods, which
are further complicated by an exponentially growing basis size and
the need to perform some kind of averaging over numerous disordered configurations.

One of the techniques that can be used to study the Physics of such
systems is Random Matrix Theory (RMT) \cite{Mehta-Book,Haake-Chaos_book,T.GuhrReview,Forrester_Book,RMT-Handbook}, which deals with the statistical
properties of eigenvalues and eigenfunctions of matrices, with random-valued elements. Since the Hamiltonians for disordered, interacting systems
involve very large matrices with random entries, the connection becomes
apparent. In recent times there have been several studies in this
direction \cite{Avishai-Richert-PRB,Modak's-paper_2014,Wigner-surmise-for-high-orde-Abul-Magd,Chavda-Kota-PLA-1,Bertrand_spin_chain_ratio_1,Iyer_Oganesyan_quasi-periodic_system,Collura_Bose_Hubbard_2}.
In this article we investigate the systematics of eigenvalue statistics
in Heisenberg spin-chains, in the presence of a random inhomogeneous
magnetic field, and the scalar-chirality three-spin interaction, whose various limits
lead to the preservation or breaking of various unitary and anti-unitary symmetries, leading
to various eigenvalue statistics which can be described by random matrix ensembles with appropriate symmetry properties, as detailed below~\cite{Dyson-3fold,Mehta-Book,T.GuhrReview,Haake-Chaos_book,Forrester_Book}.

Wigner first introduced RMT in Physics in the context of explaining the statistical
properties of nuclear spectra \cite{Wigner-before-RMT-1951,Wigner-Conference,Wigner-book-1957}. Since then RMT
has been successfully applied in the spectral study of various disordered
systems, quantum chaotic systems, and large complex atoms and molecules \cite{Mehta-1960,Mehta-Book,T.GuhrReview}.
It has also found its application in stock-market data analysis, atmospheric
science, medical science and in many more fields \cite{RMT-application-1,RMT-application-2,RMT-application-3,RMT-Handbook}.
The valuable contributions of Wigner and Dyson in explaining complex energy levels
of large nuclei \cite{Wigner-RMT-1,Wigner-RMT-2,Wigner-RMT-3,Wigner-RMT-4,Dyson-RMT}, provide deep insights into the generic properties of energy spectra
of physical many-body quantum systems. Dyson pioneered the classification of Random Matrix Ensembles via his famous \emph{threefold way} \cite{Dyson-3fold,Dyson-RMT}, depending on the presence or absence of various unitary (mainly rotations) and anti-unitary (time-reversal) symmetries in the system. The representation of most physical Hamiltonians, rendered non-integrable due to the presence of random components like disorder, involve relatively large matrices with random entries, and hence the spectral properties of such systems fall well within the ambit of RMT. Spacing distribution of energy levels in
RMT plays a decisive role as an indicator of integrability of a quantum
many-body system \cite{Integrability-1_Santos_&_Rigol,Integrability-2_Santos_&_Rigol,Integrability-3_Kollath_2010_Bose_Hubbard_1,Integrability-4_Santos,Integrability-5_Rigol_review}.
The Berry-Tabor conjecture \cite{Berry-Tabor_conjecture_paper_1977.0140}
and the BGS (Bohigas, Giannoni, and Schmit) conjecture \cite{BGS-conjecture}
asserted that the distribution of level spacings of an integrable
and a non-integrable quantum system follow the Poissonian and the Wigner-Dyson distributions, respectively. Generally, there is no
single definition of integrability \cite{Quantum-Integrability_Caux_2011}
of a quantum system (or the fact as to whether the system is exactly
solvable or not) -- one popular practice is to represent this in terms
of thermalization \cite{Rigol_From_quantum_chaos_and_eigenstate_thermalization}.
Classically, thermalization of a system is dictated by its dynamical chaotic
behavior, ergodicity, etc \cite{thermalization-1_Lebowitz_Penrose,Penrose_1979,thermalization_Giovanni}. Classical chaotic systems are non-integrable in
nature. But there is no such obvious feature which leads to the thermalization in quantum many-body systems \cite{thermalization_Srednicki,thermalization_Rigol_nature}. Quantum integrable systems are comprised of a large number of physical operators (conserved quantities) in the thermodynamic limit \cite{thermalization-2_Shastry,Integrable_model_Rigol}, that commute with the system Hamiltonian. On the other hand, quantum non-integrable systems do not
have an extensive number of conserved quantities and their dynamics is not strictly guided by conservation laws \cite{non_integrable_Deutsch,thermalization_Srednicki,non_integrable_Rigol}. 

In RMT, Nearest-neighbor spacing distribution (NNSD), \emph{viz}., the distribution
of spacings between consecutive eigenvalues, is commonly used to study the local fluctuations in the eigenspectrum. Depending upon the underlying symmetries, and the correlations among the eigenvalues,
universal features associated with Poissonian statistics, Gaussian
orthogonal ensemble (GOE) statistics, Gaussian unitary ensemble (GUE)
statistics, or Gaussian symplectic ensemble (GSE) statistics, can
follow. When the eigenvalues are uncorrelated, the NNSD follows the Poissonian distribution. Whereas
NNSD of correlated eigenvalues follows one of the three above mentioned Gaussian statistics
\cite{Dyson-RMT,Mehta_Dyson_paper,Mehta-Book}. In real physical systems, the preservation or breaking of various unitary and anti-unitary
symmetries (like time-reversal symmetry, unconventional time-reversal
symmetry, rotational symmetry, \emph{etc}.), leads the spectral fluctuations to
follow one of the above mentioned statistics. For example, the NNSD computed using the spectra of heavy nuclei \cite{Wigner-RMT-1,Wigner-RMT-2,Wigner-RMT-3,Wigner-RMT-4,Dyson-RMT}, chaotic billiards \cite{billiard_1999,Kota_Embedded_Random_Matrix_Ensembles_billiards}, non-integrable spin-chains \cite{Avishai-Richert-PRB,Modak's-paper_2014,Chavda-Kota-PLA-1}, SYK Hamiltonian \cite{SYK_Hamiltonian}, \emph{etc.}, all agree with the \emph{Wigner-surmise} results for NNSD \cite{Mehta-Book,Haake-Chaos_book,T.GuhrReview}. In addition, there can be circumstances when there is a ``\emph{partial breaking}''
of a symmetry which results in the system spectra to follow an \emph{intermediate statistics} \cite{DysonBM,Mehta-Book,Lanz-Haake-PRL,T.GuhrReview,Chavda-Kota-PLA-1,Chavda-Kota-PLA-2,Modak's-paper_2014,Modak_integrability_to_chaos,Ayana's_PhysRevE}. These systems are analyzed via crossover random matrix models or RMT interpolating functions \cite{Lanz-Haake-PRL,KumarPandey2011a,KumarPandey2011b,PandeyMehta1983,MehtaPandey1983,Forrester_Book,Modak's-paper_2014,Ayana's_PhysRevE,Schweine_NNSD_of_magnetoexcitons,Kota_Sumedha,TKS2018,Pandey1995_Brownian-motion_Model_of_Discrete_Spectra}
that depend on symmetry-breaking or RMT crossover parameters (discussed
in Sec. \ref{subsec:Level-spacing-distribution}). Limiting values
of these parameters lead to the two extremes of the concerned symmetry crossover. These RMT
crossover parameters can often be associated with certain physical parameters of the system under study, for instance
magnetic field, that breaks {\em rotational} and {\em time-reversal} symmetries~\cite{Beenakker_RMT_in_Quantum_Transport,Schweine_NNSD_of_magnetoexcitons}.

The well known \emph{Wigner-surmise} results for NNSD (discussed in
Sec. \ref{subsec:Level-spacing-distribution}) can be derived using $2\times2$ Gaussian random matrices and serve as very good approximations to the large dimension exact results~\cite{Wigner-book-1957,Mehta-1960,Mehta-Book,Wigner-Siam-review}. Over the years, among other things, these results
have been used to analyze statistical behavior of spectra from large many-body quantum systems~\cite{Mehta-Book,Wigner-surmise-for-high-orde-Abul-Magd,T.GuhrReview}.
Similar results, comprising interpolating functions, have also been empirically proposed or derived using small-dimension matrices for studying symmetry crossovers
\cite{Lanz-Haake-PRL,French_Kota_GOE-GUE,Lenz_interpolation_Poi-GOE,Caurier_1990_interpolation,Kota_Sumedha}.
These Wigner-surmise-like results have been quite fruitful in examining
NNSD crossovers in quantum many-body systems with stochastic interactions or couplings to random external fields. In quantifying the spectral fluctutations in symmetry crossovers, the physical symmetry-breaking parameters need to be suitably scaled which results in the universal validity of the Wigner-surmise-like formulae, and thereby their applicability to a large variety of systems and problems ensues \cite{Lanz-Haake-PRL,KumarPandey2011a,KumarPandey2011b,PandeyMehta1983,MehtaPandey1983,TKS2018,Pandey1981}. Some examples in this context include heavy-nuclei \cite{Pandey1988}, metallic ring in a magnetic field \cite{metallic_ring_in_mag_field}, metal-insulator transition \cite{Pragya_Shukla_metal_insulator_transition}, magneto-conductance of ballistic quantum dots \cite{magneto-conductance_ballistic_quantum_dots}, quantum kicked rotor \cite{Ayana's_PhysRevE}, bipartite entanglement entropy \cite{bipartite_entanglement_entropy}, etc. NNSD crossovers can also be utilized for studying symmetry dictated spectral crossovers in quantum spin systems~\cite{Modak's-paper_2014}.

In this work, we study the crossover between different random matrix ensembles
using linear spin-chain models similar to the ones discussed in Refs. \cite{Avishai-Richert-PRB}
and \cite{Modak's-paper_2014}. Most earlier studies, in this context, have typically used empirical or phenomenological crossover formulae. We, on the other hand, use interpolating functions based on 
crossover RMT models, such as the ones derived in Refs. \cite{Lenz_interpolation_Poi-GOE,Lanz-Haake-PRL}, where the physical meaning of the symmetry-crossover parameters is lucid.
We successfully map the crossovers in the physical spin-chain models onto the RMT crossover ensembles and thus corroborate
the associated universality. By comparing the RMT crossover parameters with the symmetry-breaking physical parameters in spin-chain models, we identify universal scaling
exponents for different NNSD crossovers in the large system size limit. In addition to NNSD, we also examine the distribution of the ratio of two consecutive
level spacings. This quantity has the advantage that, unlike the computation of NNSD, it does not require the unfolding of the spectra due
to the presence of non-uniform density of states (DOS). Due to this, the study of spacings-ratio related statistics has gained a lot of attention in recent years among researchers after being
introduced by Oganesyan and Huse in studying
statistical behavior of one-dimensional spinless-fermion model~\cite{OganesyanPRB2007}. Wigner-surmise like results for the ratio distribution (RD), based on $3\times 3$ matrix models, are available for the three standard symmetry classes of RMT, as well as for certain crossover ensembles~\cite{AtasPRL2013_ratio_1,Atas_2013_ratio_2,TKS2018,Ayana's_PhysRevE}
Over the years these results of ratio distribution have been successfully
applied in studying statistical behavior of large quantum many-body
systems like spin-chain systems \cite{Chavda-Kota-PLA-1,Chavda-Kota-PLA-2,Bertrand_spin_chain_ratio_1,Luitz_spin_chain_ratio_2,Pal_spin_chain_ratio_3,Cuevas_Spin_chain_ratio_4}, many-body quasi-periodic systems \cite{Iyer_Oganesyan_quasi-periodic_system},
Bose-Hubbard model \cite{Integrability-3_Kollath_2010_Bose_Hubbard_1,Collura_Bose_Hubbard_2}, etc. Similar to NNSD, our detailed analysis of ratio distribution in spin-chain models for various symmetry crossovers and comparison with RMT results again leads to identification of universal scaling exponents.

The presentation scheme in the rest of this article is as follows. In Sec.~\ref{sec:Spin-Hamiltonians}, we describe the spin-chain Hamiltonian used in our study and its various limits which give rise to distinct symmetry classes. In Sec.~\ref{sec:Random-Matrix-Theory} we summarize the various RMT key concepts and results used in our analysis. We present our calculations and results in Sec.~\ref{sec:Calculations-and-Results} which includes, \emph{inter alia}, a detailed explication of the universality aspects. We conclude in Sec.~\ref{sec:Conclusion} with summary and discussion of our work.

\section{Methodology : The Spin Hamiltonian, Its Symmetries, The Basis and Configuration Averaging\label{sec:Spin-Hamiltonians}}

We consider a one-dimensional spin-1/2 Hamiltonian $H$, with one spin per lattice site. It consists of the usual isotropic Heisenberg term ($H_{h}$), a random
term ($H_{r}$) representing the coupling of the spin system to a spatially inhomogeneous (on the length-scale of lattice spacings) and random magnetic field \footnote{This micro-scale inhomogeneity is essential to observe the spectral crossovers that we intend to study, because if one uses a random but spatially uniform (again on the length scale of the lattices that we use), the Zeeman term simplifies to the product of the uniform field amplitude ($h_{z}$) times the total $\mathrm{S^{z}}$ of the lattice. Such a term has no effect at all when we restrict ourselves to the $\mathrm{S^{z}}=0$ sector of a lattice with an {\em even number of sites} and the system will always remain in the Poissonian regime. Even for a lattice with an {\em odd number of sites}, for which the lowest sector is $\mathrm{S^{z}}=1/2$, such a field amounts to a uniform shift of {\em all} energy levels, that does not alter the level spacings or their ratios (as these are calculated within a given realization of the random field), in which we are interested. So again no crossovers from the Poissonian regime will be observed.}, and the 3-site scalar spin chirality term ($H_{c}$) \cite{Avishai-Richert-PRB,Modak's-paper_2014} which appears in the third order of the strong coupling expansion of the Hubbard model for a complex hopping amplitude, induced in the presence of a magnetic field \cite{Diptiman's-paper}. For example, this kind of spin correlation term can emerge while probing Mott insulators using circularly polarized laser \cite{spin_chirality_Kitamura}. We have,
\begin{align}
\nonumber
&H=H_{h}+H_{r}+H_{c}\\
&=\sum_{j=1}^{N-1}J\boldsymbol{\mathrm{S}}_{j}\cdot\boldsymbol{\mathrm{S}}_{j+1}+\sum_{j=1}^{N}h_{j}\mathrm{S^{z}}_{j}+\sum_{j=1}^{N-2}J_{t}\mathbf{S}_{j}\cdot[\mathbf{S}_{j+1}\times\mathbf{S}_{j+2}].\label{eq:totalhamil}
\end{align}
This Heisenberg spin-1/2 chain consists of $N$ sites, $\boldsymbol{\mathrm{S}}_{j}$
is the spin operator at site $j$, and $J$ is the nearest-neighbor
exchange interaction. Zeeman like effect is introduced via the random term $H_{r}$, where $h_{j}$
is the random magnetic field along $z$ direction
at site $j$, which follows a Gaussian distribution having zero mean
and variance $h^{2}$ \cite{Avishai-Richert-PRB}. The three-site interacting scalar spin-chirality term \cite{Diptiman's-paper,Spin_chirality_Freericks,Spin_chirality_Rokhsar,Spin_chirality_Wen_Wilczek}
$H_{c}$ is also included in our Hamiltonian, with a coupling constant $J_{t}$. This term can be simplified as \footnote{The $1/2$ factor in the simplified form is missing in the Refs. \cite{Avishai-Richert-PRB,Modak's-paper_2014}.
See the Appendix for derivation.},
\begin{equation}
J_{t}\mathbf{S}_{j}\cdot[\mathbf{S}_{j+1}\times\mathbf{S}_{j+2}]=\frac{1}{2}\sum_{k,l,m}iJ_{t}\varepsilon_{klm}\mathrm{S_{\mathit{k}}^{z}\mathrm{S_{\mathit{l}}^{+}\mathrm{S_{\mathit{m}}^{-}}}},\label{eq:simplified_scalar_chirality}
\end{equation}
where $\varepsilon_{klm}$ is the standard Levi-Civita symbol and
each of $k$, $l$ and $m$ takes the values $j$, $j+1$ and $j+2$ (See the Appendix for a derivation). This term favors a non-coplanar
arrangement of spins and hence counters the other two terms in this sense. All other interactions in the Hamiltonian are expressed in units of
the Heisenberg coupling $J$ throughout this paper (so effectively $J=1.0$).

It is well-known from the literature \cite{Mehta-Book,Haake-Chaos_book,Dyson-RMT,Mehta_Dyson_paper} that the eigenvalues of an integrable system are uncorrelated, therefore the eigenvalue-spacing distribution of the Hamiltonian is Poissonian. For the non-integrable systems, on the other hand, the presence or absence of various unitary and anti-unitary symmetries drive the eigenvalue-spacings of the system Hamiltonian to follow different Gaussian ensemble distributions, as described below.

Any angular momentum operator $\mathcal{J}$ is odd under the usual time-reversal
symmetry operation, i.e., $T_{0}\mathcal{J}T_{0}^{-1}=-\mathcal{J}$. The \emph{anti-unitary}
time-reversal operator for a spin-1/2 system is defined as $T_{0}=e^{i\pi\mathrm{S^{y}}/\hbar}K$
where $\mathit{\mathrm{S^{y}}}$ is the $\mathrm{y}$-component of
spin operator and $K$ is the \emph{complex conjugation} operator. The spin,
being an angular momentum, is odd under time-reversal and
for a particular term of a spin Hamiltonian, the {\em evenness} or {\em oddness} under the time-reversal symmetry
depends on the number of the spin operators involved in that term. $H_{h}$
is time-reversal symmetry invariant (or even) ($T_{0}H_{h}T_{0}^{-1}=H_{h}$),
as it involves {\em two} spin operators. But $H_{r}$
and $H_{c}$ are odd under time-reversal as the number of
the spin operators involved in these terms is odd ({\em one} and {\em three}, respectively). The \emph{unconventional time-reversal symmetry} is discussed in Refs.
\cite{Haake-Chaos_book,Avishai-Richert-PRB,Modak's-paper_2014},
where the anti-unitary unconventional time-reversal operator is defined
by $T=e^{i\pi\mathrm{S^{x}}/\hbar}T_{0}$. $H_{h}$ and $H_{r}$ are
even and $H_{c}$ is odd under the unconventional time-reversal symmetry. This is because the unitary operator $e^{i\pi\mathrm{S^{x}}/\hbar}$ reverses the signs of $\mathrm{S}_{j}^{\mathrm{y}}$ and
$\mathrm{\mathrm{S}_{j}^{z}}$ but not of $\mathrm{S}_{j}^{\mathrm{x}}$. So, in totality, $T$ reverses the sign of $\mathrm{S}_{j}^{\mathrm{x}}$ alone. Thus, the full Hamiltonian $H$ is neither even nor odd under these anti-unitary symmetry operations.

Now, when only the Heisenberg term is present in Eq. (\ref{eq:totalhamil}),
the Hamiltonian of the system is preserved under both anti-unitary
symmetries, $T_{0}$ and $T$. So the eigenvalues of the Hamiltonian matrix are uncorrelated
and the level-spacings are expected to follow the Poissonian distribution. If we
switch on $H_{r}$ then $T_{0}$ is broken but the Hamiltonian $H_{1}$ ($=H_{h}+H_{r}$) is still invariant under $T$. The Hamiltonian matrix becomes real-symmetric
and the correlation statistics of the eigenvalues should follow the GOE
distribution. When all the three terms of Eq. (\ref{eq:totalhamil})
are present, both anti-unitary symmetries are broken and the matrix
representation of $H$ becomes complex Hermitian (complex nature
is introduced by the scalar-chirality term, apparent from Eq. (\ref{eq:simplified_scalar_chirality})).
The correlation statistics of the eigenvalues should now follow GUE
distribution. By varying the relative amplitudes of each terms in
$H$, we can achieve crossovers between the Poissonian, GOE and GUE
distributions.

To construct the basis for our exact diagonalization calculations, we consider a spin-chain system
with a spin-1/2 at each lattice site. That means each of the $N$
sites of the system can be occupied by either of an up-spin ($\mathrm{m^{z}}=1/2\equiv$ $\uparrow$) or a down-spin ($\mathrm{m^{z}}=-1/2\equiv$ $\downarrow$),
where $\mathrm{m^{z}}$ is the eigenvalue of $\mathrm{S^{z}}.$ There
exists $2^{N}$ number of basis states ($\left|\mathrm{m_{1}^{z}m_{2}^{z}m_{3}^{z}\mathbf{....}m_{\mathit{N}}^{z}}\right\rangle $)
for an $N$-site system. Here we can see that $\left[\mathrm{S}^{2},H\right]\neq0$
(where $\mathrm{S^{2}}$ is the square of the total spin operator
with $\mathrm{\mathbf{S}}=\mathop{\sum_{j=1}^{N}\mathbf{S}_{j}}$)
but $\left[\mathrm{S^{z}},H\right]=0$ (where the total $\mathrm{z}$-component
of spin $\mathrm{S^{z}}=\mathrm{\sum_{j=1}^{\mathit{N}}}\mathrm{S}_{j}^{\mathrm{z}}$).
So among the total $2^{N}$ number of basis states we only consider
a subspace having a constant $\mathrm{S^{z}}$ value. In our calculations
we have considered systems where $N$ is \emph{even} and we restrict ourselves to the $\mathrm{S^{z}}=0$ subspace, which includes contributions from
all total $\mathrm{S}$ sectors and has the largest dimension ($n=C_{N/2}^{N}$).
As we are considering the whole spectrum, the sign of $J$ is irrelevant unlike in ground state studies. We have used full exact diagonalization methods to obtain the energy eigenvalues and eigenfunctions, as we need the full spectrum of eigenvalues. This consequently limits the system sizes that we can access, to some extent, but manages to capture the universality behavior that we demonstrate in this work.

Since we are considering the system in the presence of a random, inhomogeneous magnetic field, one typically needs to average over several random configurations or realizations of the magnetic field, in order to obtain good spectral statistics. In a real-life experimental situation, this could be due to the coupling of the system to an external magnetic field, fluctuating rapidly on the time-scale of a typical magnetic measurement. This means that any measured physical quantity is essentially time-averaged over several configurations of this rapidly fluctuating magnetic field, which is mimicked by an ensemble average over several random configurations in our calculations, implicitly implying usual statistical {\em ergodicity} \cite{thermalization-1_Lebowitz_Penrose,Penrose_1979,thermalization_Giovanni}. Each \emph{configuration}, for a given $h$ (standard deviation of the Gaussian distribution from which the magnetic fields are drawn), consists of a set of $N$ site magnetic fields $\{h_{j}\}$. For each such set, the Hamiltonian is diagonalized in the above basis and the energy spacings are computed and stored. For the purposes of configuration averaging, we consider a set of $\mathcal{M}$ such configurations, henceforth referred to as an \emph{ensemble}. This diagonalization and computation of spacings, is repeated for each member of the ensemble, and the final set of all energy spacings from all the configurations is used to plot the histogram (distribution) of energy spacings. As $h$ is varied over some suitable grid, the entire process above is repeated and a level-spacing distribution is obtained for each $h$. Details of the exact number of configurations used in an ensemble, for each lattice size, are discussed in the Sec.~\ref{sec:Calculations-and-Results}. We find that, the larger the lattice size (and hence the basis size), the smaller the number of configurations are needed to obtain good statistics, or equivalently, a smooth distribution profile. This indeed reflects the "self-averaging", or the "spectral ergodicity" property of RMT, which demonstrates the equivalence of the ensemble average and the spectral average, in the limit of large Hamiltonian matrices \cite{Haake-Chaos_book,Pandey1979_ergodicity}. This also applies to the spectral studies of disordered many-body quantum systems for subspaces with relatively large basis sizes \cite{Santos_self_averaging_1,Santos_self_averaging_2}.

\section{Random Matrix Theory (RMT)\label{sec:Random-Matrix-Theory}}

In this section, we summarize the RMT results which are required for the analysis of our results. These include the exact expressions for the nearest neighbor spacing distributions and ratio distributions, for invariant as well as crossover ensembles.

\subsection{Nearest-Neighbor Spacing Distribution (NNSD) \label{subsec:Level-spacing-distribution}}

In RMT, the fluctuation properties of a system's eigenspectrum is quantified
using both short-range and the long-range level statistics~\cite{Mehta-Book,Haake-Chaos_book}.
NNSD is the most widely studied statistical measure to quantify
the local-correlations among energy eigenvalues of a given system. Since the DOS is nonuniform for the eigenspectrum of an arbitrary physical system, one requires to implement the unfolding procedure to compare the calculated fluctuations with the standard RMT results. In our analysis pertaining to NNSD, we use polynomial fits to the calculated DOS and use it to unfold the spectra~\cite{Mehta-Book,Haake-Chaos_book,Avishai-Richert-PRB,Schweine_NNSD_of_magnetoexcitons}.
If the ordered sequence of $n$ energy eigenvalues of a Hamiltonian
is given by $\mathop{\varepsilon_{1}<\cdots<\varepsilon_{n},}$
then the unfolded eigenvalues are calculated using $\tilde{\varepsilon}_{j}=\int_{\varepsilon_{1}}^{\varepsilon_{j}} \rho(\varepsilon)d\varepsilon$, where $\rho(\varepsilon)=d\mathcal{N}(\varepsilon)/d\varepsilon$ is the fitted DOS, and $\mathcal{N}(\varepsilon)$ is the cumulative DOS. The nearest-neighbor level spacing of the unfolded eigenvalues is defined as $s_{j}=\tilde{\varepsilon}_{j+1}-\tilde{\varepsilon}_{j}$. The corresponding frequency distribution, or rather its fitted envelope ($P(s)$), is then compared with the RMT analytical results for its quantification, as discussed in the introduction. The Wigner surmise expressions for the three Dyson symmetry classes along with the Poisson distribution are compiled in Table \ref{tab:NNSD formulas}, for this purpose.

\begin{table}
\caption{Probability distribution of nearest-neighbor spacings for unfolded
eigenvalues \cite{Mehta-Book,Haake-Chaos_book}.\label{tab:NNSD formulas}}
\begin{tabular}{cc}
\toprule 
{\footnotesize{}Type of distribution} & {\footnotesize{}NNSD Probability Density}\tabularnewline
\midrule
{\footnotesize{} Poissonian} & $P_{Poi}(s)=\exp(-s)$\tabularnewline
{\footnotesize{}GOE} & {\footnotesize{}$\mathop{P_{GOE}(s)=(\pi s/2)\exp(-\pi s^{2}/4)}$}\tabularnewline
{\footnotesize{}GUE} & {\footnotesize{}$\mathop{P_{GUE}(s)=(32s^{2}/\pi^{2})\exp(-4s^{2}/\pi)}$}\tabularnewline
\bottomrule
\end{tabular}
\end{table}

In this paper, we are interested in studying crossover between different RMT ensembles. It is modeled using the Pandey-Mehta Hamiltonian~\cite{PandeyMehta1983,MehtaPandey1983,Lenz_Haake_crossover_matrix},
\begin{equation}
\label{eq:crossover_random_matrix}
\mathcal{H}=\frac{\mathcal{H}_{0}+\lambda\mathcal{H}_{\infty}}{\sqrt{1+\lambda^{2}}},
\end{equation}
 wherein the symmetry of the primary Hamiltonian $\mathcal{H}_{0}$
is broken by the perturbing Hamiltonian $\mathcal{H}_{\infty}$, with the extent of this symmetry breaking being controlled by the crossover-parameter $\lambda$. The two extremes of $\mathcal{H}$, \emph{viz.}, $\mathcal{H}_{0}$ and $\mathcal{H}_{\infty}$, belonging to two different symmetry classes, are obtained for $\lambda=0$ and $\lambda\to\infty$. Our focus in the current work is on examining the Poissonian to GOE and GOE to GUE crossovers, which is achieved by tuning the crossover parameter. The Wigner-surmise like results for the spacing distributions have been derived in Ref. \cite{Lanz-Haake-PRL} for these crossovers. The NNSD function interpolating between the Poissonian and the GOE distributions is given by \cite{Lanz-Haake-PRL,Schierenberg_surmise_like_crossover},
\begin{align}
\label{eq:interpolate_POI to GOE}
\nonumber
\mathit{f_{\mathrm{\mathit{P}}\rightarrow O}}(\lambda,s)=\frac{su^{2}(\lambda)}{\lambda}\exp\left(-\frac{u^{2}(\lambda)s^{2}}{4\lambda^{2}}\right)\\
\times\intop_{0}^{\infty}\exp(-x^{2}-2x\lambda)I_{0}\left(\frac{sxu(\lambda)}{\lambda}\right)dx.
\end{align}
Here $I_{0}(z)$ is the modified Bessel function and $u(\lambda)=\sqrt{\pi}U(-1/2,0,\lambda^{2})$, with $U(a,b,z)$ being the confluent Hypergeometric function of the second kind (Tricomi function). The limiting
cases of $\lambda\rightarrow0$ and $\lambda\rightarrow\infty$, lead to the Poissonian and the GOE spacing distributions, respectively, as shown in Table~\ref{tab:NNSD formulas}. Similarly, the interpolating function
for GOE to GUE crossover is given by \cite{Lanz-Haake-PRL,Schierenberg_surmise_like_crossover},
\begin{align}
\label{eq:interpolate_GOE_TO_GUE}
\nonumber
 & f_{\mathrm{\mathit{O}}\rightarrow U}(\lambda,s)=\left(\frac{2+\lambda^{2}}{2}\right)^{1/2}D^2(\lambda) \\
 & \times s \exp\left(\frac{s^{2}D^{2}(\lambda)}{2}\right)\mathrm{erf}\left(\frac{sD(\lambda)}{\lambda}\right),
\end{align}
where 
\[
D(\lambda)=\left(\frac{\pi(2+\lambda^{2})}{4}\right)^{1/2}\left[ 1\text{\textminus}\frac{2}{\pi}\left(\mathrm{arctan}\left(\frac{\lambda}{\sqrt{2}}\right)\text{\textminus}\frac{\sqrt{2}\lambda}{2+\lambda^{2}}\right)\right] .
\]
In this case, the limits $\lambda \to0$ and $\lambda \to \infty$ result in the GOE and the GUE spacing distributions, respectively, as given in Table~\ref{tab:NNSD formulas}. These interpolating functions are based on $2\times2$ matrix models, however, once the transition parameter is suitably scaled, these results can be applied to larger dimensional cases also
\cite{Schweine_NNSD_of_magnetoexcitons,Brody_surmise_like_crossover,Berry_surmise_like_crossover,Schierenberg_surmise_like_crossover,Schweiner_surmise_like_crossover,Kota_Sumedha,Izrailev_surmise_like_crossover,Hasegawa_surmise_like_crossover,Casati_surmise_like_crossover,Abul_Magd_surmise_like_crossover}.
One of our key objectives in this paper is to probe such a scaling for spin-chain systems.

\subsection{Ratio Distribution (RD) \label{subsec:Ratio-distribution}}

As discussed in the introduction, another short-range fluctuation measure which circumvents the need of unfolding and has become quite popular in recent times is the distribution of the ratios of consecutive level spacings~\cite{OganesyanPRB2007,Chavda-Kota-PLA-1,Chavda-Kota-PLA-2,AtasPRL2013_ratio_1,Atas_2013_ratio_2,Armando_Angel_Distribution_of_the_ratio_of_consecutive_level_spacings,Bertrand_spin_chain_ratio_1}. Considering again the ordered sequence of $n$ energy eigenvalues $\mathop{\varepsilon_{1}<\cdots<\varepsilon_{n}}$, one examines the distribution of the ratio of two consecutive level spacings given by
$r_{j}=(\varepsilon_{j+2}-\varepsilon_{j+1})/(\varepsilon_{j+1}-\varepsilon_{j})$. The RMT results for the ratio distribution, $\mathcal{P}(r)$, based on $3\times 3$ matrix models, have been derived for the three standard symmetry classes, along with the Poissonian case in Ref.~\cite{AtasPRL2013_ratio_1}. These are analogous to the results for NNSD and are compiled in Table \ref{tab:RD formulas}. A related quantity is $\tilde{r}=\min(r,1/r)$, which is also used quite often as this quantity gets restricted in the interval $[0,1]$, and the corresponding probability distribution is $\tilde{\mathcal{P}}(\tilde{r})=2 \mathcal{P}(\tilde{r})\Theta(1-\tilde{r})$.

\begin{table}
\caption{Probability distribution of ratio of consecutive spacings~\cite{AtasPRL2013_ratio_1,Atas_2013_ratio_2}.\label{tab:RD formulas}}
\begin{tabular}{cc}
\toprule 
{\footnotesize{}Type of distribution} & {\footnotesize{}RD Probability Density}\tabularnewline
\midrule
{\footnotesize{} Poissonian} & $\mathcal{P}_{Poi}(r)=1/(1+r)^2$\tabularnewline
{\footnotesize{}GOE} & {\footnotesize{}$\mathop{\mathcal{P}_{GOE}(r)=27r(r+1)/[8(r^2+r+1)^{5/2}]}$}\tabularnewline
{\footnotesize{}GUE} & {\footnotesize{}$\mathop{\mathcal{P}_{GUE}(r)=81\sqrt{3}r^2(r+1)^2}/[4\pi(r^2+r+1)^4]$}\tabularnewline
\bottomrule
\end{tabular}
\end{table}

As far as the crossover is concerned, there are several phenomenological formulae available for the RD, see for example Refs.~\cite{Chavda-Kota-PLA-1,Chavda-Kota-PLA-2,Armando_Angel_Distribution_of_the_ratio_of_consecutive_level_spacings}. However, we are interested in results based on RMT crossover model as in Eq.~\eqref{eq:crossover_random_matrix}. In this context, to the best of our knowledge, there is no result available for the Poissonian to GOE crossover and we resort to numerics. On the other hand, an exact expression for the ratio distribution, based on a $3\times 3$ matrix model, is known for the GOE to GUE crossover~\cite{Ayana's_PhysRevE}.
It is given by the following interpolating function
\begin{align}
\nonumber
\mathcal{F}_{\mathrm{\mathit{O}}\rightarrow U}(\lambda,r)=\frac{r(1+r)}{16\sqrt{6}\pi}(1+\lambda^{2})^{3/2}\\
\times\left[g(a,b)+g(a,br)-g(a,br+b)\right],\label{eq:interpolate_ratio_GOE-GUE}
\end{align}
 where 
\begin{equation*}
g(\eta,\xi)=\frac{\xi(5\eta^{2}+3\xi^{2})}{\eta^{4}(\eta^{2}+\xi^{2})^{2}}+\frac{3}{\eta^{5}}\arctan\left(\frac{\xi}{\eta}\right),
\end{equation*}
 $a(r)=\sqrt{(1+r+r^{2})/6}$ and $b(\lambda)=1/(2\sqrt{2}\lambda)$.
When $\lambda\rightarrow0$ this interpolating function follows GOE
statistics and for $\lambda\rightarrow\infty$ it follows GUE statistics, as given in Table~\ref{tab:RD formulas}.
Similar to the case of NNSD, when applied to large-dimensional matrices, the crossover parameter $\lambda$ in the above expressions needs to be scaled properly as we will see in Sec. \ref{subsec:Ratio-distribution:}.

\section{Calculations and Results\label{sec:Calculations-and-Results}}

In this section, we present the details of our calculations and various results concerning RMT analysis of the spin-chain system for the various symmetry crossovers.

\subsection{Nearest-Neighbor Spacing Distribution (NNSD)\label{subsec:Nearest-Neighbor-Spacing-Distrib}}

We consider nearest-neighbor spacings of energy eigenvalues computed 
from the Hamiltonian matrix $H$ by its numerical diagonalization and analyze its statistics. We notice that, level spacing distributions for spin-chain systems with lattice size $N=14$ (matrix dimension for the $\mathrm{S^{z}}=0$ subspace, $n=3432$) and above, follow standard RMT results \footnote{We have studied the level spacing statistics for smaller systems having $N=10$ ($n=252$) and $N=12$ ($n=924$), they do not follow the standard RMT ensembles, closely enough.}. So, we carry out calculations for spin-chains having $N=14$ and $N=16$ ($n=12870$) sites, on an ensemble of $\mathcal{M}=50$ and $10$ matrices, respectively.  Also, performing diagonalization
for systems having $N=18$ ($n=48620$) or more, is computationally
very expensive, so we do not attempt the full basis calculation, but resort to a reduced basis calculation, based on spin-inversion symmetry consideration, as discussed later. Periodic boundary conditions give rise to extra degeneracies in the spectra and therefore we consider
open boundary condition throughout our calculations. We have
unfolded the system spectra using polynomial fits (as mentioned in Sec. \ref{subsec:Level-spacing-distribution})
to obtain unfolded spectra for which the average spacing equals unity.

\subsubsection{Poissonian to GOE Crossover \label{subsec:Poissonian-to-GOE}}

\begin{figure}[h]
\includegraphics[scale=0.25]{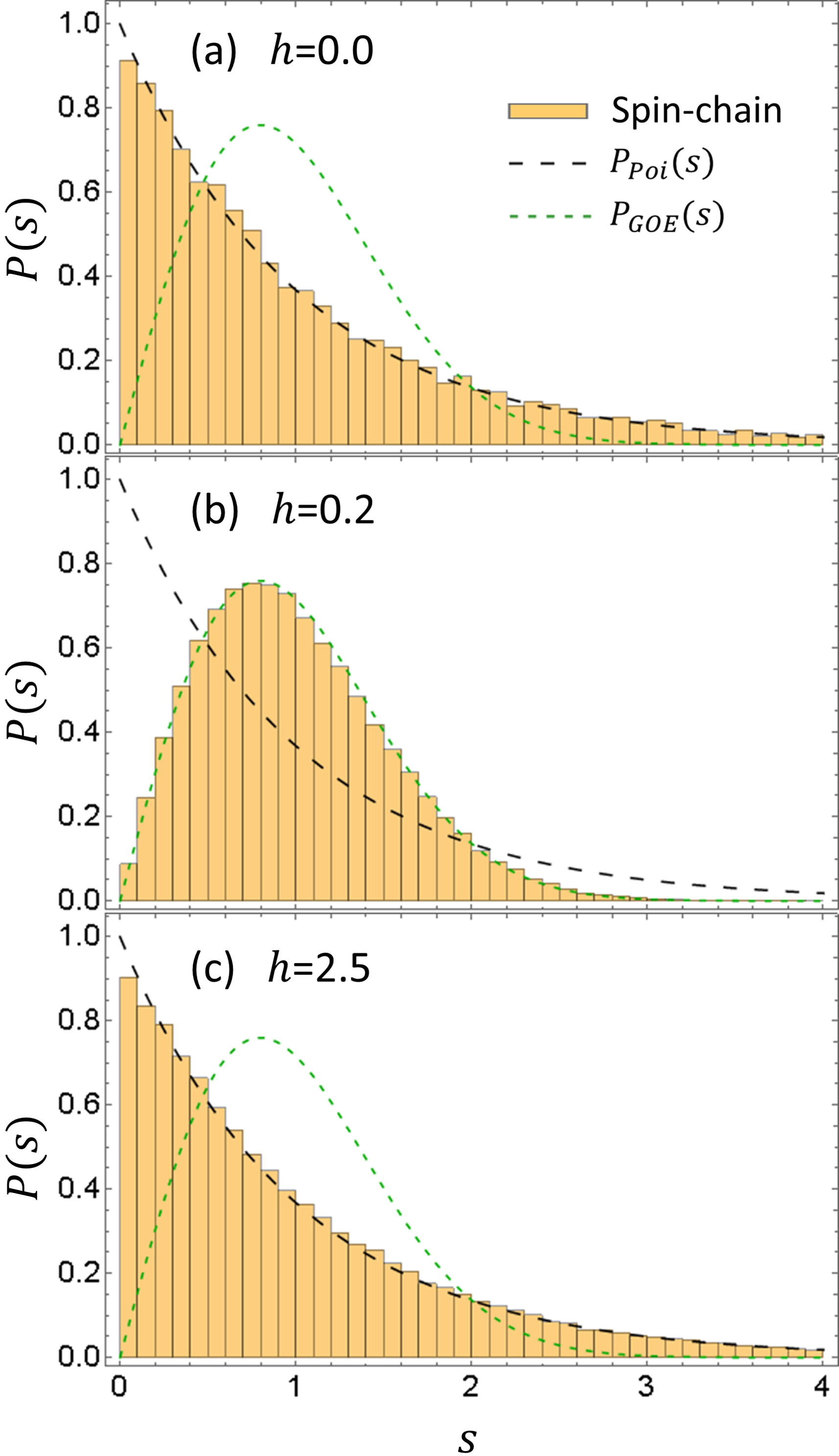}

\caption{NNSD for $N=16;$ (a) shows Poissonian distribution when magnetic
field is zero; (b) NNSD follows GOE for $h=0.2$ and (c) shows how Poisson distribution is recovered due to eigenvector localization for a typical large
magnetic field ($h=2.5$).\label{fig:Poi-GOE-Poi}}
\end{figure}

To realize Poissonian to GOE crossover we proceed as follows. To begin with, we switch off the scalar chirality
term ($J_{t}=0$) and the random field term ($h=0$). In this setting, the system is
integrable and NNSD obeys Poissonian level spacing distribution ($P_{Poi}(s)$), as expected.
We then switch on the random term and vary the magnetic field values $h_j$ on the sites. As mentioned earlier, these are drawn from the Gaussian distribution with mean zero and standard deviation $h$. This introduces
disorder in the system and the system becomes non-integrable.  Each set, $\{h_{j}\}$, of the $N$ random site fields corresponds to one member of the aforesaid ensemble or one realization of disorder, as already discussed earlier. As the magnitude of $h$ is gradually increased, a crossover from Poissonian to GOE statistics is observed. After a certain value of $h$, the regime beyond the applicability of RMT is reached and the statistics reverts back to Poissonian, due to \emph{localization of eigenstates}. This simply means that, due to a very high magnetic field, the quantum fluctuation effects included in the Heisenberg term are strongly suppressed, and consequently the weights of the eigenstates spread over a much fewer number of basis configurations now. In Fig. \ref{fig:Poi-GOE-Poi} we show this
Poissonian$\rightarrow$GOE$\rightarrow$Poissonian (integrable$\rightarrow$quantum chaotic$\rightarrow$integrable) transition with the increasing random magnetic field value.

\begin{figure}
\includegraphics[scale=0.18]{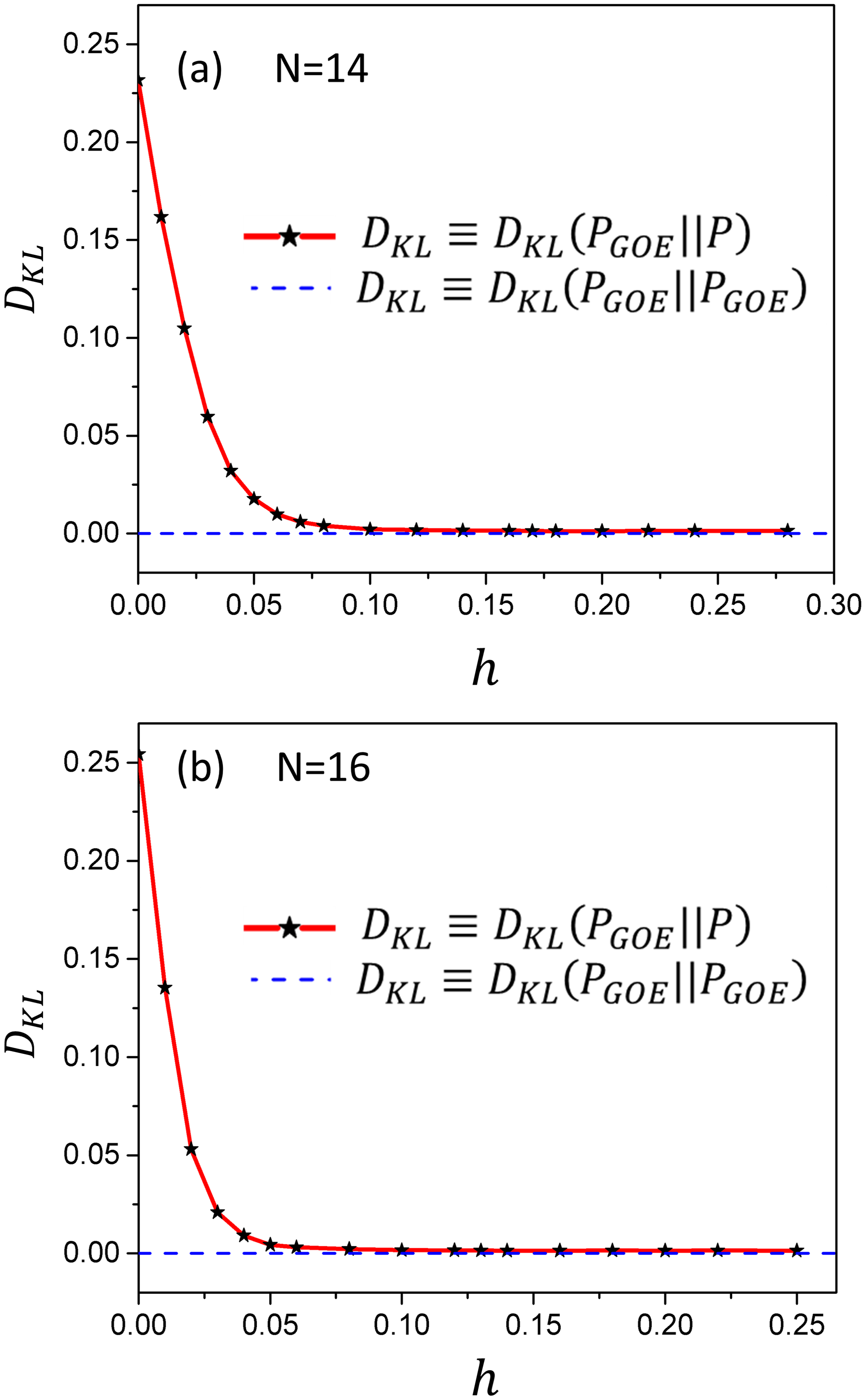}

\caption{Variation of $D_{KL}(P_{GOE}||P)$ (KL divergence of NNSD $P(s)$
with respect to $P_{GOE}(s)$) with increasing value of the symmetry
breaking crossover parameter $h$ for the (a) $N=14$ and (b) $N=16$
systems. \label{fig:KL_div_Poi_to_GOE}}
\end{figure}

To estimate the threshold for the magnetic field at which the Poisson-GOE crossover
is more or less complete, we evaluate the Kullback-Leibler divergence (KL divergence) \cite{kullback_Leibler_1951,Kullback_KL_divergence_1953,kullback1968-book}
of the calculated spacing distribution $P(s)$ at different values of $h$, with respect to $P_{GOE}(s)$. This comparison is shown in Fig. \ref{fig:KL_div_Poi_to_GOE} for the $N=14$ and $N=16$ systems. The
KL divergence is a measure of the difference between two
probability distributions. Let us say $p$ and $q$ are the two probability
distributions in question, then the KL divergence of
$q$ relative to $p$ is given by,
\begin{equation}
\label{eq:KL-Divergence_formula}
D_\text{KL}(p||q)=\int dx\, p(x)\log\left(\frac{p(x)}{q(x)}\right).
\end{equation} 
For our computation, we use the discretized version of this formula.
When the two distributions are same, i.e., $p=q,$ the KL divergence
is \emph{zero}, and for very distinct distributions, it assumes a high value. Here, the two extreme cases correspond to Poissonian and GOE distributions for which KL divergence has a value $\sim0.28$ (using the analytical forms $P_{Poi}(s)$ and $P_{GOE}(s)$, from Table \ref{tab:NNSD formulas}). In Fig. \ref{fig:KL_div_Poi_to_GOE}, we observe that these extreme KL divergence values are $0.24$ and $0.26$ for the $N=14$ and $N=16$ systems, respectively, converging towards the aforesaid theoretical value of $\sim0.28$.

From Fig. \ref{fig:KL_div_Poi_to_GOE}
we estimate that the Poissonian to GOE crossover is achieved at about $h=0.17$ and $h=0.13$ for the $N=14$ and $N=16$ systems, respectively.
Beyond the crossover value of $h$, the systems enter in the GOE regime
and KL divergence saturates near zero as observed in Fig. \ref{fig:KL_div_Poi_to_GOE} .

\begin{figure}
\includegraphics[scale=0.25]{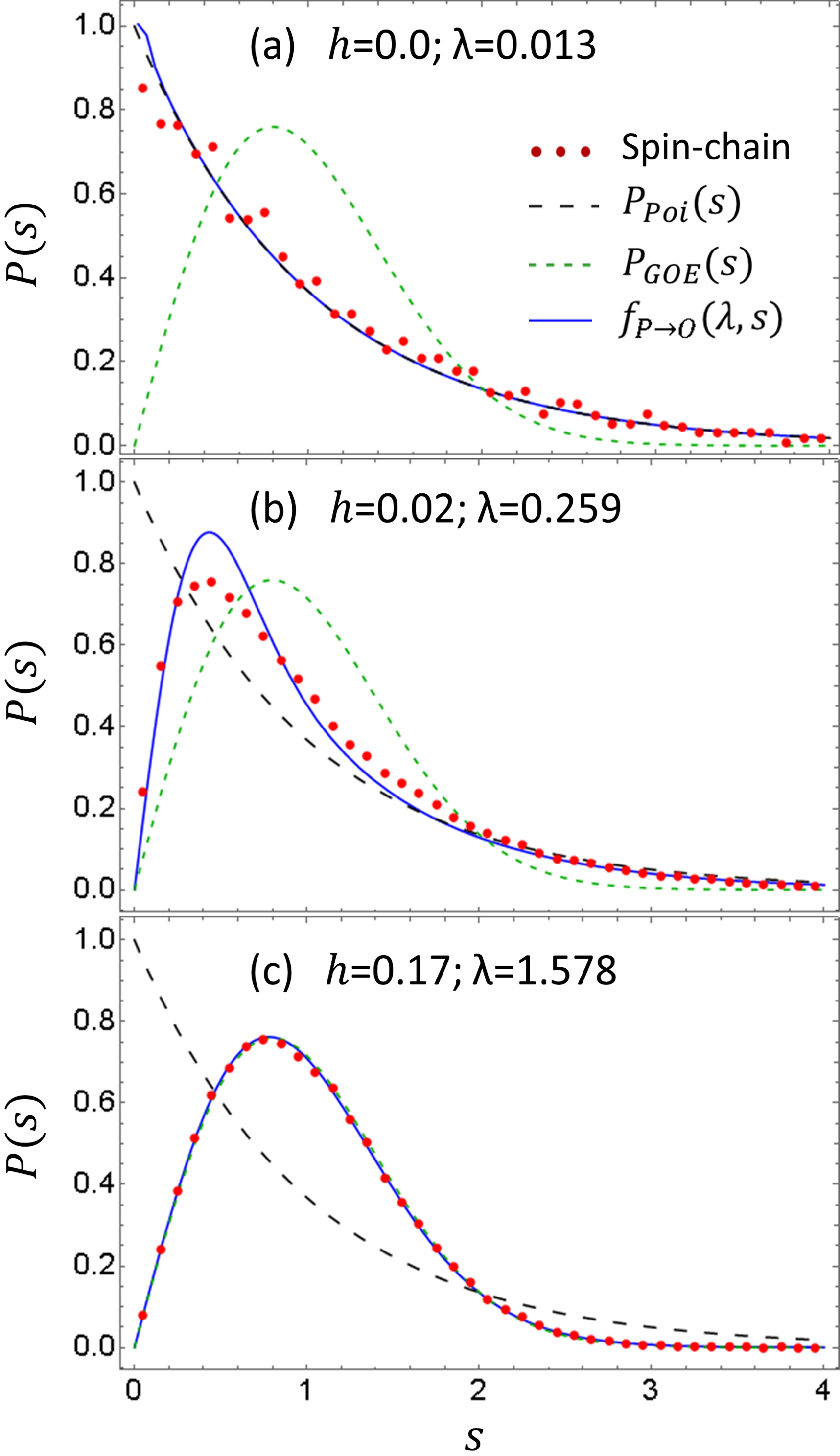}

\caption{NNSD for $N=14$; Poissonian to GOE crossover with increasing $h$.
(a) and (c) show the two limiting cases Possonian and GOE respectively
and (b) shows one of the intermediate cases. We have used $\mathit{f_{\mathrm{\mathit{P}}\rightarrow O}}(\lambda,s)$
to fit this crossover and the values of the RMT crossover parameter
$\lambda$ are determined.\label{fig:Poi-GOE for N=00003D14}}
\end{figure}

\begin{figure}
\includegraphics[scale=0.25]{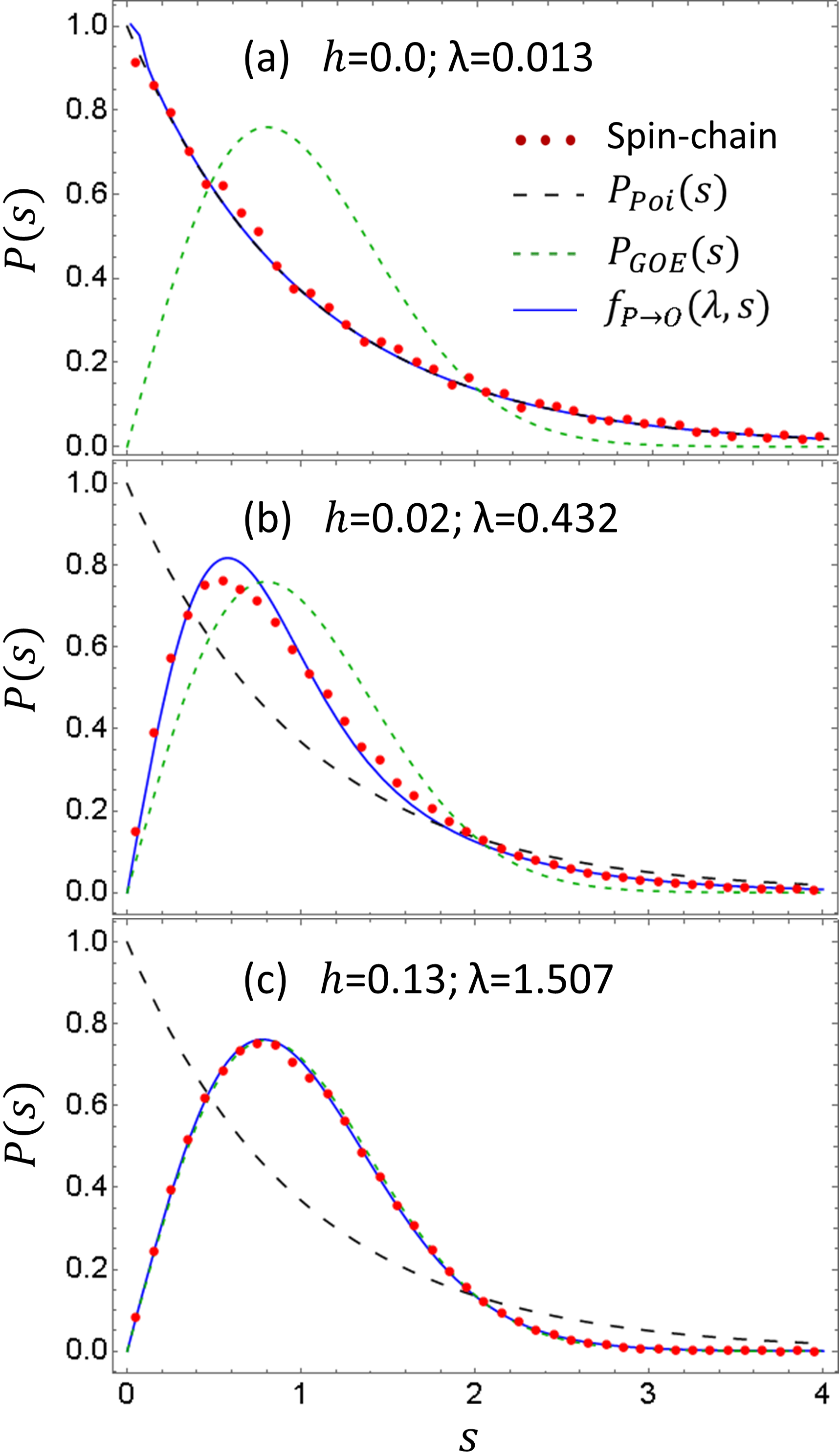}

\caption{\label{fig:Poi-GOE for N=00003D16}Plots of NNSD for $N=16$,
similar to the $N=14$ cases in Fig. \ref{fig:Poi-GOE for N=00003D14}.}
\end{figure}

Next, we plot the NNSD for the Poissonian to GOE crossover for $N=14$ (Fig.
\ref{fig:Poi-GOE for N=00003D14}) and $N=16$ (Fig. \ref{fig:Poi-GOE for N=00003D16})
systems, and find out that both systems reach GOE at external magnetic
field values which are estimated  from the KL divergence calculation as $h\sim0.17$ and $h\sim0.13$, respectively. We observe here
that {\em with increasing system size we need a lesser magnitude of the symmetry breaking random magnetic field}, which is in agreement with RMT crossover studies in Refs. \cite{Pandey1981,PandeyMehta1983,KumarPandey2011b}. We now fit the RMT interpolating function $\mathit{f_{\mathrm{\mathit{P}}\rightarrow O}}(\lambda,s)$ of Eq. (\ref{eq:interpolate_POI to GOE}) to these computed distributions to obtain the values of the RMT crossover parameter $\lambda$, vis-a-vis the physical crossover parameter $h$. We find that the best fits for $N=14$, the above two extremes are obtained respectively for $\lambda=0.013$ and $1.578$, whereas for $N=16$, the respective values are $0.013$ and $1.507$. Figs. \ref{fig:Poi-GOE for N=00003D14} and \ref{fig:Poi-GOE for N=00003D16} show the corresponding plots.

\subsubsection{GOE to GUE Crossover \label{subsec:GOE-to-GUE}}

\begin{figure}
\includegraphics[scale=0.18]{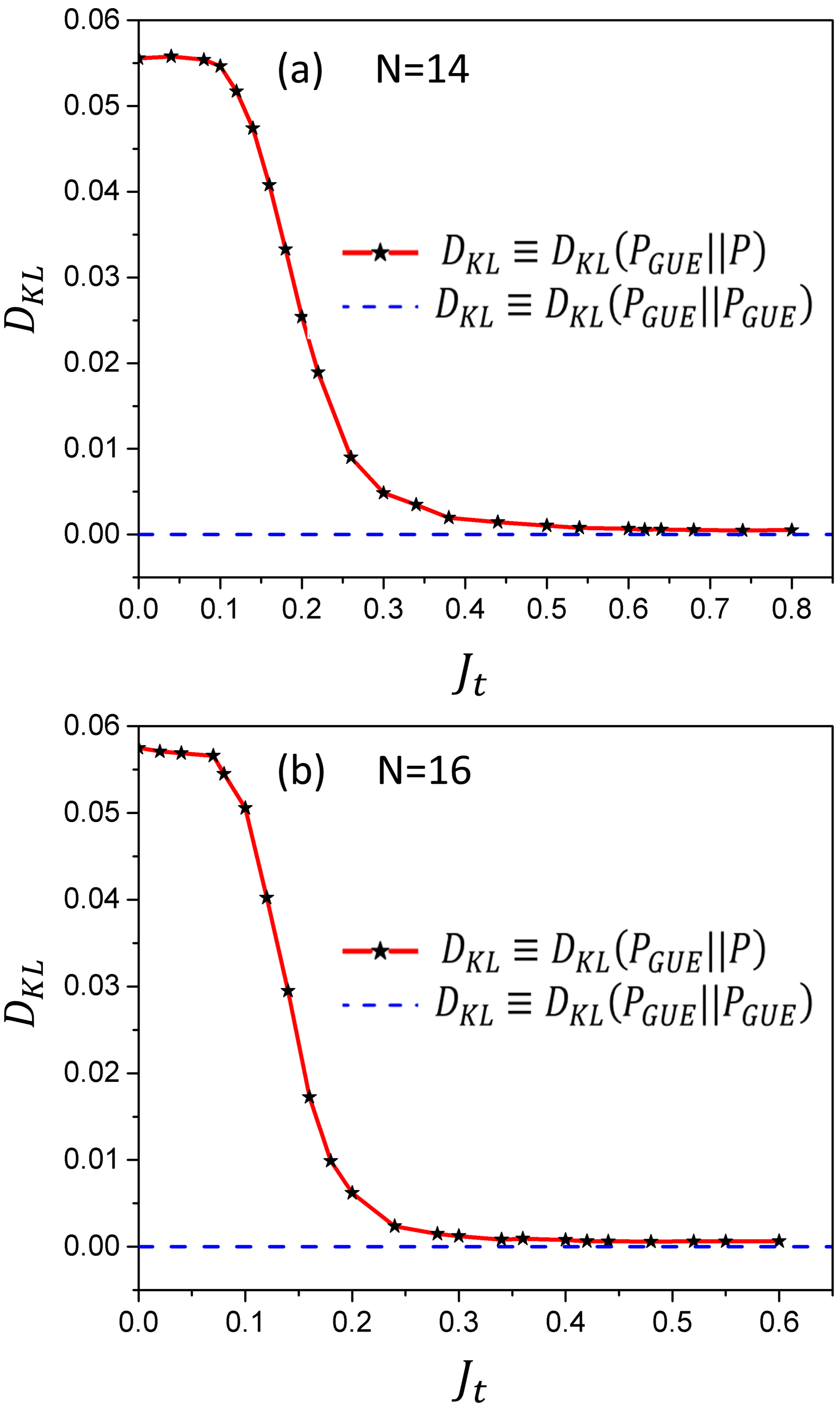}
\caption{Variation of $D_{KL}(P_{GUE}||P)$ (KL divergence of NNSD $P(s)$
with respect to $P_{GUE}(s)$) with the increasing value of symmetry
breaking crossover parameter $J_{t}$ for (a) $N=14$ and (b) $N=16$
systems. \label{fig:KL_div_GOE_to_GUE}}
\end{figure}

To realize GOE to GUE crossover we switch on the scalar chirality term
in Eq. (\ref{eq:totalhamil}) by keeping the system in GOE regime,
for which the value of $h$ is fixed at $0.2$ (see Figs. \ref{fig:Poi-GOE-Poi}(b) and \ref{fig:KL_div_Poi_to_GOE}). We can observe GOE to GUE crossover by slowly increasing
the amplitude of scalar chirality term $J_{t}$. Again, to estimate
the exact crossover points from GOE to GUE distributions, we have used the KL divergence of $P(s)$ with respect to $P_{GUE}(s)$ and plotted
$D_\mathrm{KL}$ against the symmetry breaking physical crossover parameter $J_{t}$
in Fig. \ref{fig:KL_div_GOE_to_GUE}. Here, the two extreme cases correspond to GOE and GUE distributions for which KL divergence has a value $\sim0.047$ (using the analytical forms $P_{GOE}(s)$ and $P_{GUE}(s)$, from Table~\ref{tab:NNSD formulas}). We observe in Fig. \ref{fig:KL_div_GOE_to_GUE} that these extreme KL divergence values are $0.055$ and $0.057$ for $N=14$ and $N=16$ systems, respectively. While increasing $J_{t}$ from zero, unlike the Poissonian
to GOE crossover case, here we observe that some threshold amount
of scalar spin-chirality amplitude ($J_{t}$) is needed to even start breaking the
GOE symmetry (as indicated by the plateauing at low $J_{t}$), and further increment of $J_{t}$ shows a linear variation
with $\lambda$. For $N=14$ and $N=16$ systems, these threshold
values are $(J_{t})_\mathrm{Th}\sim0.1$ and $(J_{t})_\mathrm{Th}\sim0.07$ respectively.
The KL divergence saturates at around $J_{t}=0.62$ and
at around $J_{t}=0.42$ for the $N=14$ and $N=16$ systems respectively. So we claim that for these values the GOE to GUE crossover is complete. 
As we have kept $h$ fixed at $0.2$, we again conclude that
with increasing system size, we can break the GOE symmetry of
the system with lesser magnitude of scalar spin chirality and
achieve GUE the distribution sooner.

\begin{figure}
\includegraphics[scale=0.25]{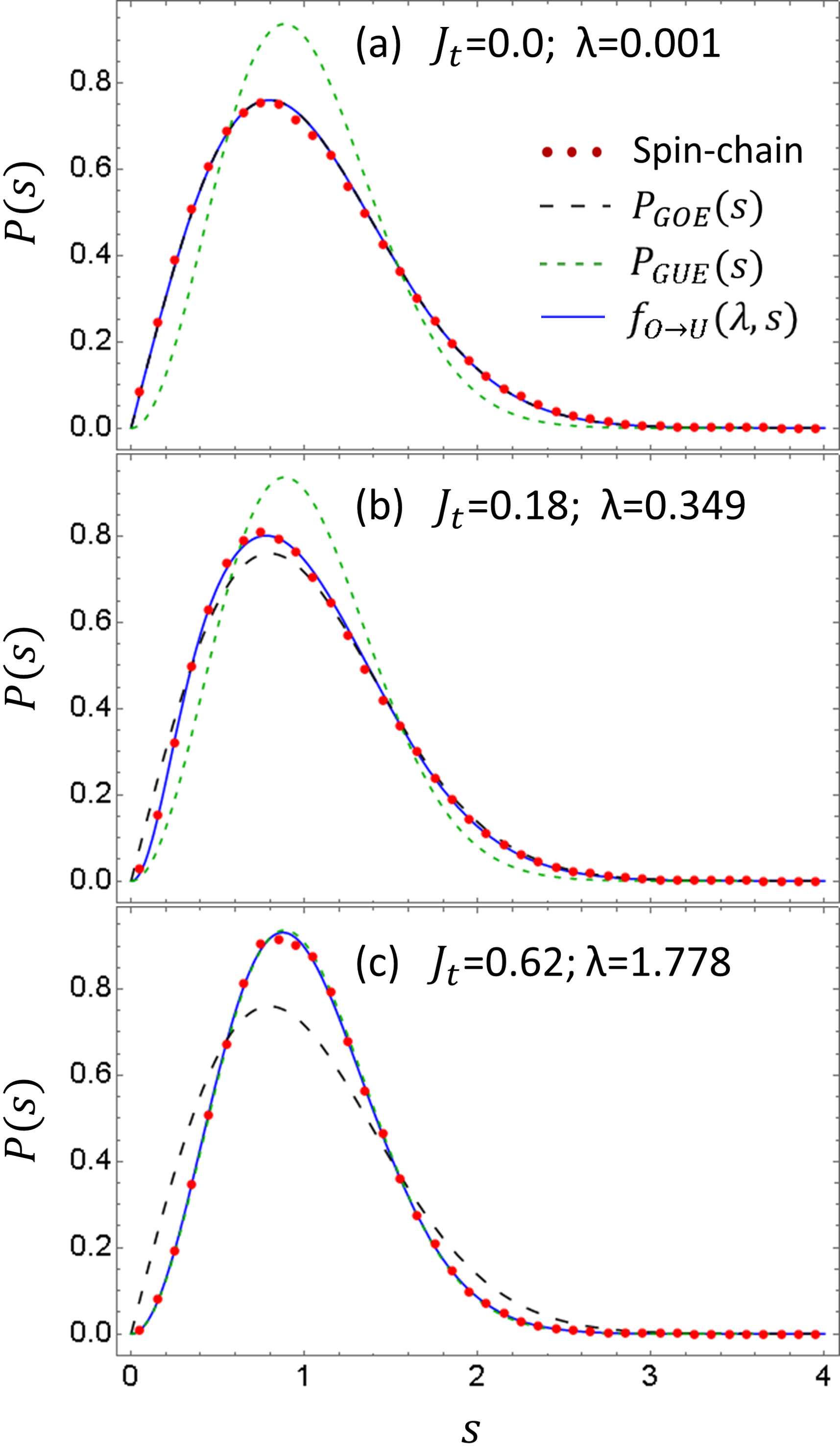}

\caption{\label{fig:GOE-GUE for N=00003D14}NNSD for $N=14$; GOE to GUE crossover
with increasing scalar spin-chirality amplitude ($J_{t}$) by keeping
the system at constant random magnetic field ($h=0.2$). (a) and
(c) show the two limiting cases GOE and GUE respectively and (b) shows
one of the intermediate states. We have used $\mathit{f_{\mathrm{\mathit{O}}\rightarrow U}}(\lambda,s)$
to fit this crossover and the value of the RMT crossover parameter
$\lambda$ is determined.}
\end{figure}

\begin{figure}
\includegraphics[scale=0.25]{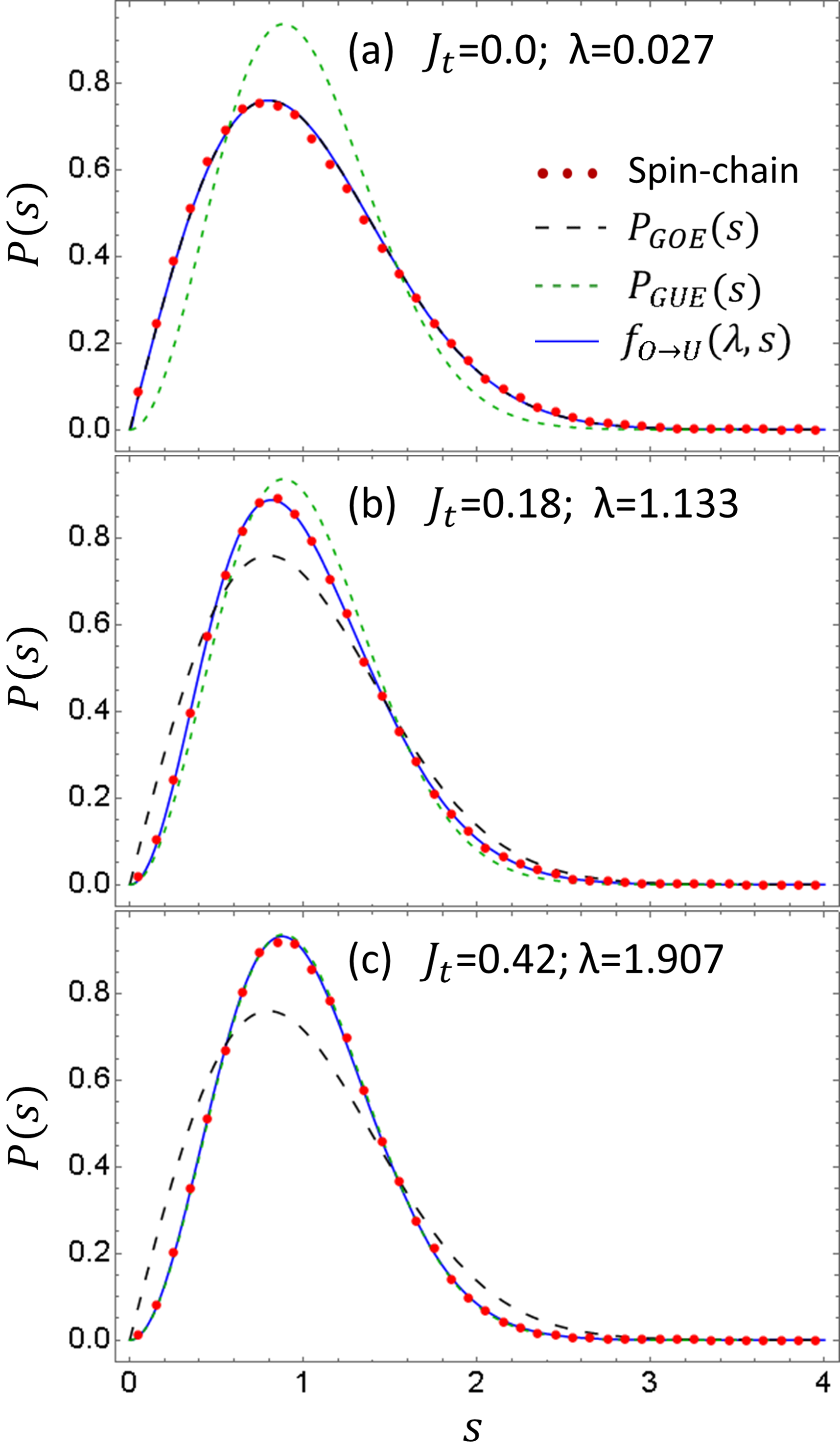}

\caption{\label{fig:GOE-GUE for N=00003D16}Plots of NNSD for $N=16$,
similar to the $N=14$ cases in Fig. \ref{fig:GOE-GUE for N=00003D14}.}
\end{figure}

Now, the GOE to GUE computed crossover data is mapped on to the interpolating function $\mathit{f_{\mathrm{\mathit{O}}\rightarrow U}}(\lambda,s)$ of Eq. (\ref{eq:interpolate_GOE_TO_GUE}) and the values of RMT crossover
parameter $\lambda$ is noted against the physical crossover parameter
$J_{t}$. We observe from Fig. \ref{fig:GOE-GUE for N=00003D14} that in the GOE regime, the value of the
RMT crossover parameter $\lambda$ is very close to zero, and at around
$\lambda\thicksim1.7$, $\mathit{f_{\mathrm{\mathit{O}}\rightarrow U}}(\lambda,s)$
matches quite well with GUE distribution, for the $N=14$ system (Fig. \ref{fig:GOE-GUE for N=00003D14}), and for the $N=16$ system it achieves the GUE distribution for the RMT crossover value of $\lambda=1.9$ (see Fig. \ref{fig:GOE-GUE for N=00003D16}, for details).

\subsubsection{Crossovers using Inverted Spin Basis \label{subsec:Inverted-spin-basis:}}

\begin{figure}
\includegraphics[scale=0.19]{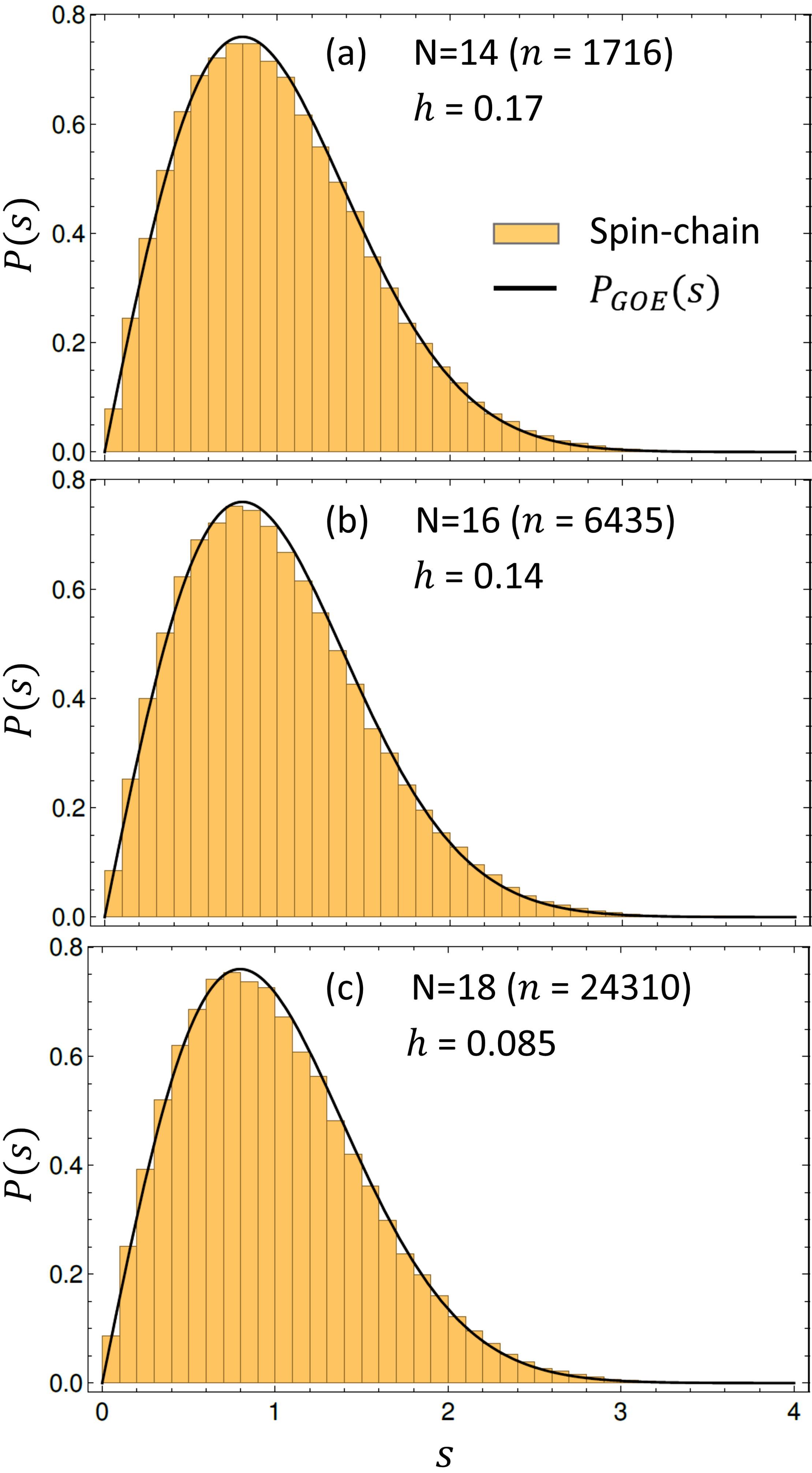}

\caption{Inverted basis GOE distribution for the (a) $N=14$, (b) $N=16$ and (c)
$N=18$ systems, at the crossover points.\label{fig:Inverted-basis-GOE}}

\end{figure}

\begin{figure}
\includegraphics[scale=0.19]{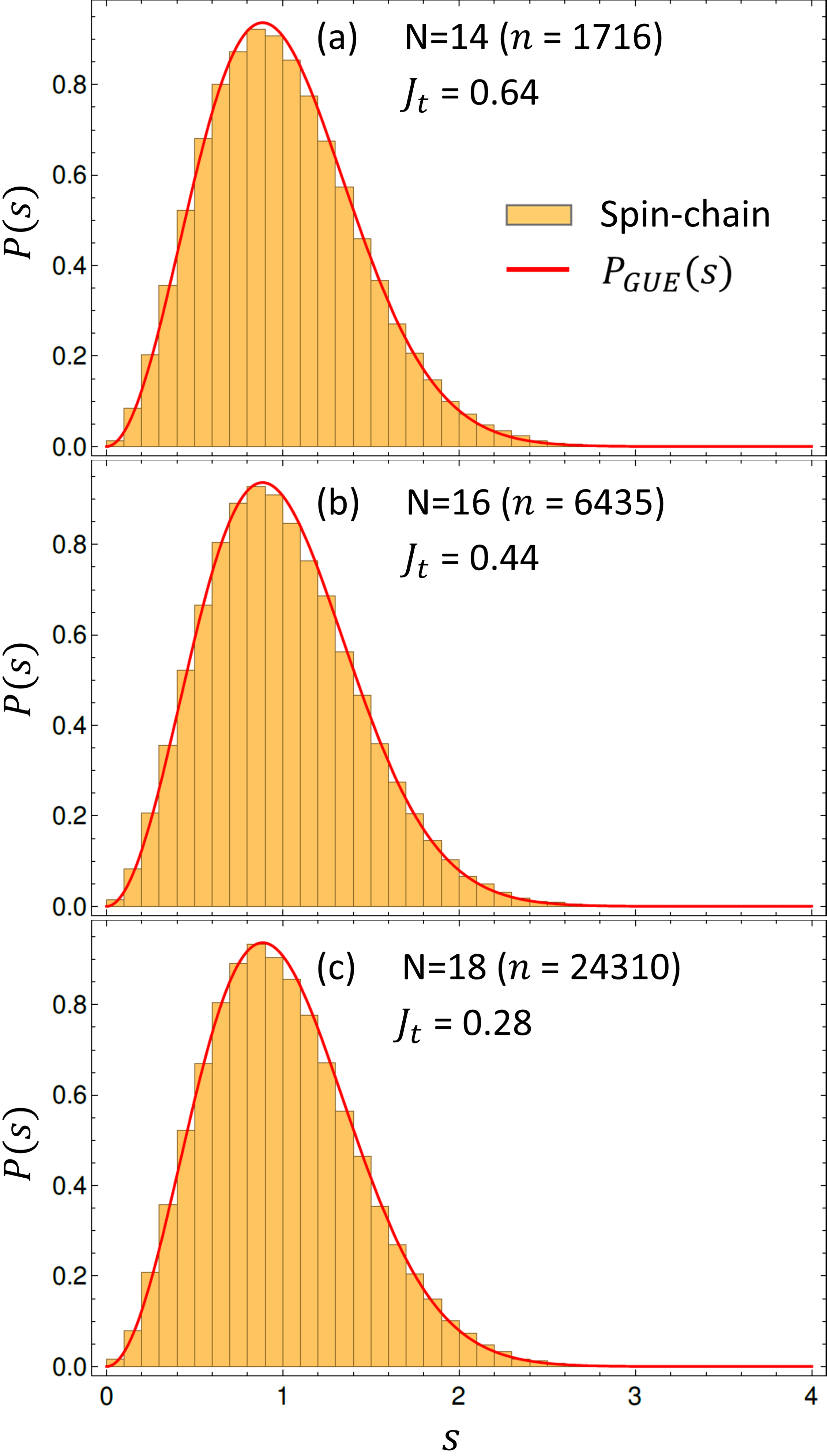}

\caption{Inverted basis GUE distribution for the (a) $N=14$, (b) $N=16$ and (c)
$N=18$ systems, at the crossover points.\label{fig:Inverted-basis-GUE}}

\end{figure}

In this work, we are also interested in exploring the universality behavior of RMT spectral crossovers for our spin-chain systems, in the large system size limit. For this, we need to explore more than just two lattice sizes. This would also enable us to properly examine the existence of a universal scaling exponent (discussed in the next section). However, we have already pointed out the computational limitations to carry out calculations for systems beyond the lattice size $N=16$. To circumvent this problem, we resort to a symmetry consideration to further reduce our system basis, as discussed below \cite{Santos_Gubin_Inverted_spin}. In our calculations we have used spin-chains having even number of
sites and as discussed in detail earlier, restrict our calculations to the $\mathrm{S^{z}}=0$ subspace. In this
scenario, there exists pair of equivalent basis vectors which are
related via global spin-inversion \cite{Santos_Gubin_Inverted_spin}.
If $\left|\downarrow\downarrow\uparrow\uparrow\downarrow\uparrow\right\rangle $
denotes a basis of $6$-site system having three up-spin and and three
down-spin, then the equivalent pair basis is denoted by the basis
$\left|\uparrow\uparrow\downarrow\downarrow\uparrow\downarrow\right\rangle $.
Now if we choose one basis vector out of each pair, then we have two
subsets of basis vectors in the $\mathrm{S^{z}}=0$ subspace. We want to
carry out NNSD calculations considering only one such subset and compare
with the full basis results. In Figs. \ref{fig:Inverted-basis-GOE}(a)
and \ref{fig:Inverted-basis-GOE}(b) we have plotted NNSD of
$N=14$ ($n=1716$) and $N=16$ ($n=6435$) systems, considering $\mathcal{M}=100$ and 20, respectively.
We observe that the Poissonian to GOE crossovers take place around the same symmetry
breaking $h$ values when compared to the full basis calculations in
Sec. \ref{subsec:Poissonian-to-GOE}. We have also studied GOE to GUE crossovers
in Figs. \ref{fig:Inverted-basis-GUE}(a) and \ref{fig:Inverted-basis-GUE}(b)
for the $N=14$ and $N=16$ systems, respectively, and observed that the values of the physical crossover parameter, $J_{t}$, obtained from the inverted basis calculations are
in good agreement with the results obtained from full basis calculations
in Sec. \ref{subsec:GOE-to-GUE}. The $N=18$ system full basis ($n=48620$)
calculations are not possible due to computational limitations. However, in view of the above observations, we study the $N=18$ system using inverted basis ($n=24310$). For this, we consider $\mathcal{M}=5$ and obtain the spacing distributions. These results are depicted in Figs. \ref{fig:Inverted-basis-GOE}(c) and \ref{fig:Inverted-basis-GUE}(c), where
we find the Poissonian to GOE crossover value for the $N=18$ system is the $h=0.085$, and the GOE to GUE crossover takes place at around $J_{t}=0.28$, for the same system, respectively. In the next
sections, along with the $N=14$ and $16$ systems with $\mathrm{S^{z}}=0$ basis, we use the $N=18$ system with inverted spin basis, in all ratio distribution calculations
and in universality studies.

\begin{figure}
\includegraphics[scale=0.2]{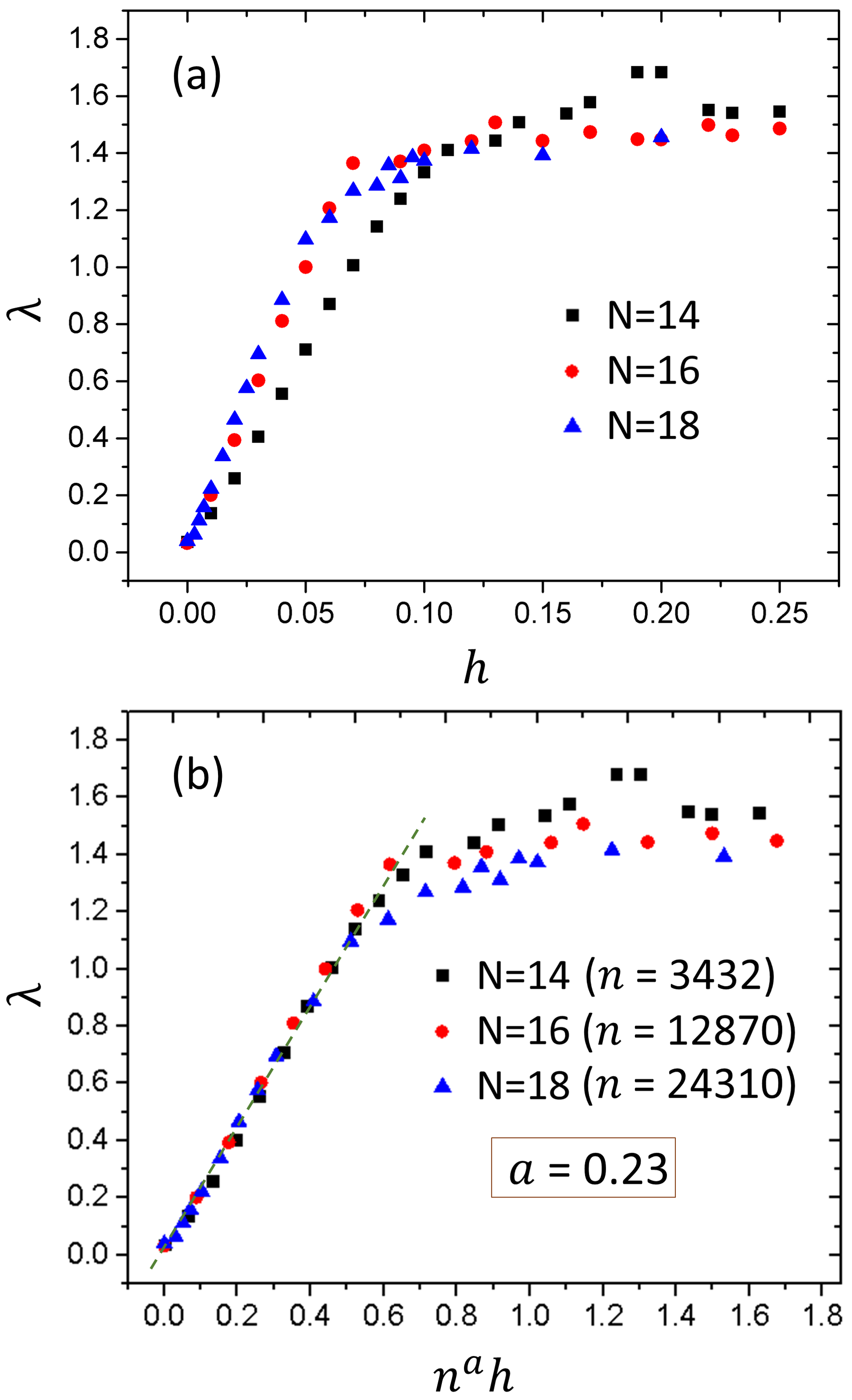}

\caption{\label{fig:scaling Poi-GOE_NNSD}Variation of RMT crossover parameter
$\lambda$ (of the crossover function $\mathit{f_{\mathrm{\mathit{P}}\rightarrow O}}(\lambda,s)$)
with, (a) physical crossover parameter $h$, for the $N=14$, $N=16$ and $N=18$ (inverted basis) systems. Poissonian to GOE crossover is observed when $\lambda$ gets saturated,
for $N=14$ the crossover value is $h=0.17$, for $N=16$ it is
$h=0.13$, and for $N=18$ it is $h=0.085$. (b) The linear region of the plots are fitted with the
scaling function $n^{a}h$. $\lambda$ is plotted against $n^{a}h$.
The universal scaling exponent $a=0.23.$
The dashed green line is a fit based on the data points occurring
in the linear regime. }
\end{figure}

\subsubsection{Universality in NNSD \label{subsec:Universality in NNSD}}

In the preceding subsections, we used the RMT interpolation functions based on $2\times 2$ matrix models to fit the calculated data and obtained the crossover parameter $\lambda$. In order to investigate the universal aspects of RMT spectral crossovers, as discussed in the Introduction, we show that after suitable scaling of the physical parameters $h$ and $J_t$, the crossovers are governed by identical values of the $\lambda$-parameter for different lattice sizes, thereby establishing universality of the RMT crossover results.
For the Poissonian to GOE crossover we plot the variation of the RMT interpolating
parameter $\lambda$ (of the crossover function $\mathit{f_{\mathrm{\mathit{P}}\rightarrow O}}(\lambda,s)$) against the physical crossover parameter $h$ in Fig. \ref{fig:scaling Poi-GOE_NNSD}(a).
Initially it varies linearly with increasing magnetic field value,
then it saturates as it approaches the GOE regime. We fit the linear
part of this plot using a scaling function $n^{a}h$, where $n$ is
the dimension of the system Hamiltonian ($n=3432$ for $N=14$, $n=12870$ for $N=16$, and $n=24310$ for $N=18$ (inverted basis)) and $a$ is the scaling exponent.
In Fig. \ref{fig:scaling Poi-GOE_NNSD}(b) $\lambda$ is plotted
against $n^{a}h$, the universal scaling exponent is found to be $a=0.23$. Therefore, we conclude
that we have found universality in the Poissonian
to GOE crossover, in the large system size limit. 

\begin{figure}
\includegraphics[scale=0.2]{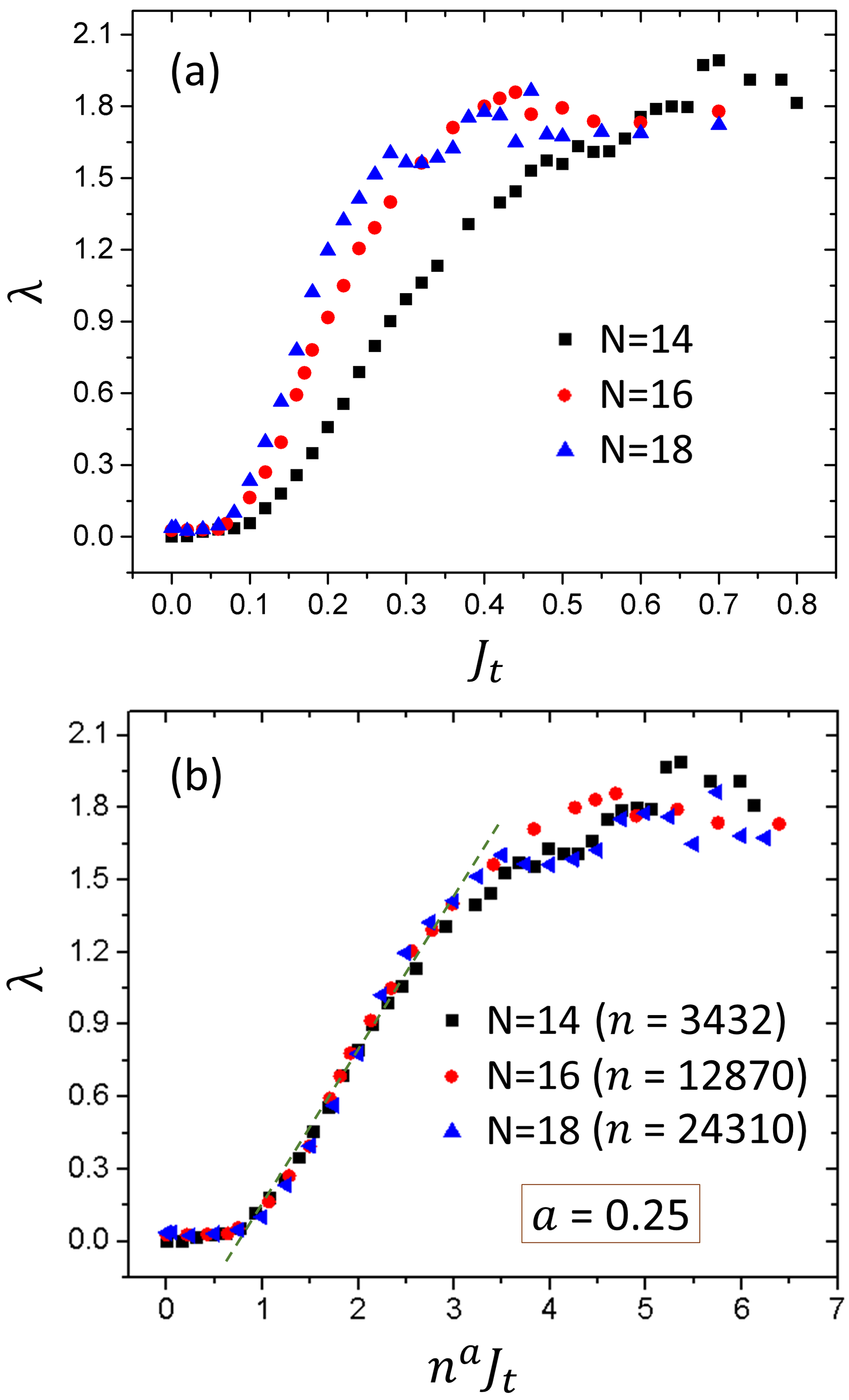}

\caption{\label{fig:scaling GOE-GUE_NNSD}Variation of RMT crossover parameter
$\lambda$ (of the crossover function $f_{\mathrm{\mathit{O}}\rightarrow U}(\lambda,s)$)
with, (a) physical crossover parameter $J_{t}$ for $N=14$, $N=16$, and
$N=18$ (inverted basis) systems. GOE to GUE crossover is observed when $\lambda$ gets saturated,
for $N=14$ the crossover value is $J_{t}=0.62$, for $N=16$ it is $J_{t}=0.42$, and for $N=18$ it
is $J_{t}=0.28$. (b) The linear region of the plots are fitted with
the scaling function $n^{a}J_{t}$. $\lambda$ is plotted against
$n^{a}J_{t}$. The universal scaling exponent $a=0.25.$ The dashed green line is a fit based on the data
points occurring in the linear regime. }
\end{figure}

Similar behavior can be observed for the GOE to GUE crossover
in Fig. \ref{fig:scaling GOE-GUE_NNSD}(a), where we have plotted
the variation of RMT interpolating parameter $\lambda$ (of the crossover
function $f_{\mathrm{\mathit{O}}\rightarrow U}(\lambda,s)$) against
the physical crossover parameter $J_{t}$. Again we have fitted the
linear increment part in Fig. \ref{fig:scaling GOE-GUE_NNSD}(b)
using the scaling function $n^{a}J_{t}$ and find that the universal
scaling exponent $a$ takes the value $0.25$ (using $N=14$, $N=16$ and
$N=18$ (inverted basis) systems). After the linear region, further increment of $J_{t}$ saturates the
crossover parameter $\lambda$ into the GUE region. Thus, similar to the Poissonian to GOE crossover, here also we observe universality in the GOE $\rightarrow$ GUE crossover, in the large system size limit.

\subsection{Ratio Distribution (RD) \label{subsec:Ratio-distribution:}}

We now examine the spacings-ratio distribution. To estimate the threshold of physical crossover parameters for which spectral crossovers are achieved, we have computed the KL divergence for NNSD. Similarly, we have studied the KL divergence for RD and find that the $D_{KL}$ vs. physical crossover parameter variations are not much different from NNSD. However, in the present case, it is interesting to study the average of the ratios, $\left\langle r\right\rangle $
and $\left\langle \tilde{r}\right\rangle $, which also help classify the system in one regime or the other. It should be noted that such a study is not useful in the case of NNSD, since the unfolding of eigenspectra necessarily makes the average spacing unity. The averages $\left\langle r\right\rangle =\intop_{0}^{\infty}r\mathcal{P}(r)dr$,
and $\left\langle \tilde{r}\right\rangle =\intop_{0}^{1}\tilde{r}\mathcal{\tilde{P}}(\tilde{r})d\tilde{r}$
can be computed and the exact RMT values of these average ratios
are included in Table \ref{tab:Theoretical-and-calculated-avg-ratios}. We have calculated $\left\langle r\right\rangle $ and $\left\langle \tilde{r}\right\rangle $
for both the Poissonian to GOE, and the GOE to GUE crossovers considering
the $\mathord{N=14,16}$ and $\mathord{18}$ systems. We have listed our calculated values of $\left\langle r\right\rangle$ and $\left\langle \tilde{r}\right\rangle$
at RMT crossovers, for different ensembles in Table \ref{tab:Theoretical-and-calculated-avg-ratios}. 

\begin{table}[t]
{\footnotesize{}\caption{Exact RMT values (from the Refs. \cite{AtasPRL2013_ratio_1,Armando_Angel_Distribution_of_the_ratio_of_consecutive_level_spacings}) and the calculated (this work) values of $\left\langle r\right\rangle $
and $\left\langle \tilde{r}\right\rangle $. \label{tab:Theoretical-and-calculated-avg-ratios}}
}{\footnotesize\par}

{\footnotesize{}}%
\begin{tabular}{>{\centering}p{1.2cm}>{\centering}m{2.2cm}>{\centering}m{2.2cm}>{\centering}m{2.2cm}}
\hline 
Quantity & Poissonian & GOE & GUE\tabularnewline
\hline 
$\left\langle r\right\rangle _{RMT}$ & $\infty$ & $\mathord{1.75}$

$1.80$ & $1.3607$

$1.3758$\tabularnewline
\hline 
$\left\langle r\right\rangle$ & $7.080$$\,$$(N=14)$

$11.41$$\,$$(N=16)$

$10.40$$\,$$(N=18)$ & $1.772$$\,$$(N=14)$

$1.757$$\,$$(N=16)$

$1.755$$\,$$(N=18)$ & $1.371$$\,$$(N=14)$

$1.372$$\,$$(N=16)$

$1.362$$\,$$(N=18)$\tabularnewline
\hline 
$\left\langle \tilde{r}\right\rangle _{RMT}$ & $0.3863$ & $0.5359$

$0.5278$ & $0.6026$

$0.5977$\tabularnewline
\hline 
$\left\langle \tilde{r}\right\rangle$ & $0.395$$\,$$(N=14)$

$0.385$$\,$$(N=16)$

$0.382$$\,$$(N=18)$ & $0.531$$\,$$(N=14)$

$0.531$$\,$$(N=16)$

$0.530$$\,$$(N=18)$ & $0.599$$\,$$(N=14)$

$0.599$$\,$$(N=16)$

$0.597$$\,$$(N=18)$\tabularnewline
\hline 
\end{tabular}{\footnotesize\par}
\end{table}

At Poissonian limit, the exact RMT value, $\left\langle r\right\rangle _{RMT}$
is $\mathord{\infty}$ \cite{AtasPRL2013_ratio_1,Armando_Angel_Distribution_of_the_ratio_of_consecutive_level_spacings},
but in practice we get relatively large finite values for our systems. The exact RMT value, $\left\langle \tilde{r}\right\rangle _{RMT}$ is $0.386$ at Poissonian
distribution and it is well in agreement with the calculated
$\left\langle \tilde{r}\right\rangle$ values for our systems. We find that, $\mathord{\left\langle r\right\rangle }$
achieves the values $\mathord{1.772}$ ($h=0.17$), $\mathord{1.757}$
$(h=0.13)$, and $\mathord{1.755}$ $(h=0.085)$ in the GOE regime for $N=14,16$
and $18$ systems, respectively, against the RMT values listed
in Table \ref{tab:Theoretical-and-calculated-avg-ratios}. Please see the Supplemental Material for a detailed variation of $\mathord{\left\langle r\right\rangle }$ and $\left\langle \tilde{r}\right\rangle$ with $h$ and $J_t$ \footnote{See Supplemental Material at [URL will be inserted by publisher] for tabulations of average ratios with varying crossover parameters $h$ and $J_t$.}. Our systems
saturate to the GUE regime at around $\mathord{\left\langle r\right\rangle }=1.37$
($\left\langle \tilde{r}\right\rangle =0.599$) for $N=14,16$ and
at $1.36$ ($0.597$) for $N=18$. Our spin-chain model conforms to different RMT symmetry classes with the variation
of physical crossover parameters $\mathord{h}$ and $\mathord{J_{t}}$,
and the calculated values of $\mathord{\left\langle r\right\rangle }$
and $\mathord{\left\langle \tilde{r}\right\rangle }$ are in good
agreement with the theoretical RMT values $\left\langle r\right\rangle _{RMT}$ and $\left\langle \tilde{r}\right\rangle _{RMT}$ \cite{AtasPRL2013_ratio_1,Armando_Angel_Distribution_of_the_ratio_of_consecutive_level_spacings}.

\begin{figure}[H]
\includegraphics[scale=0.2]{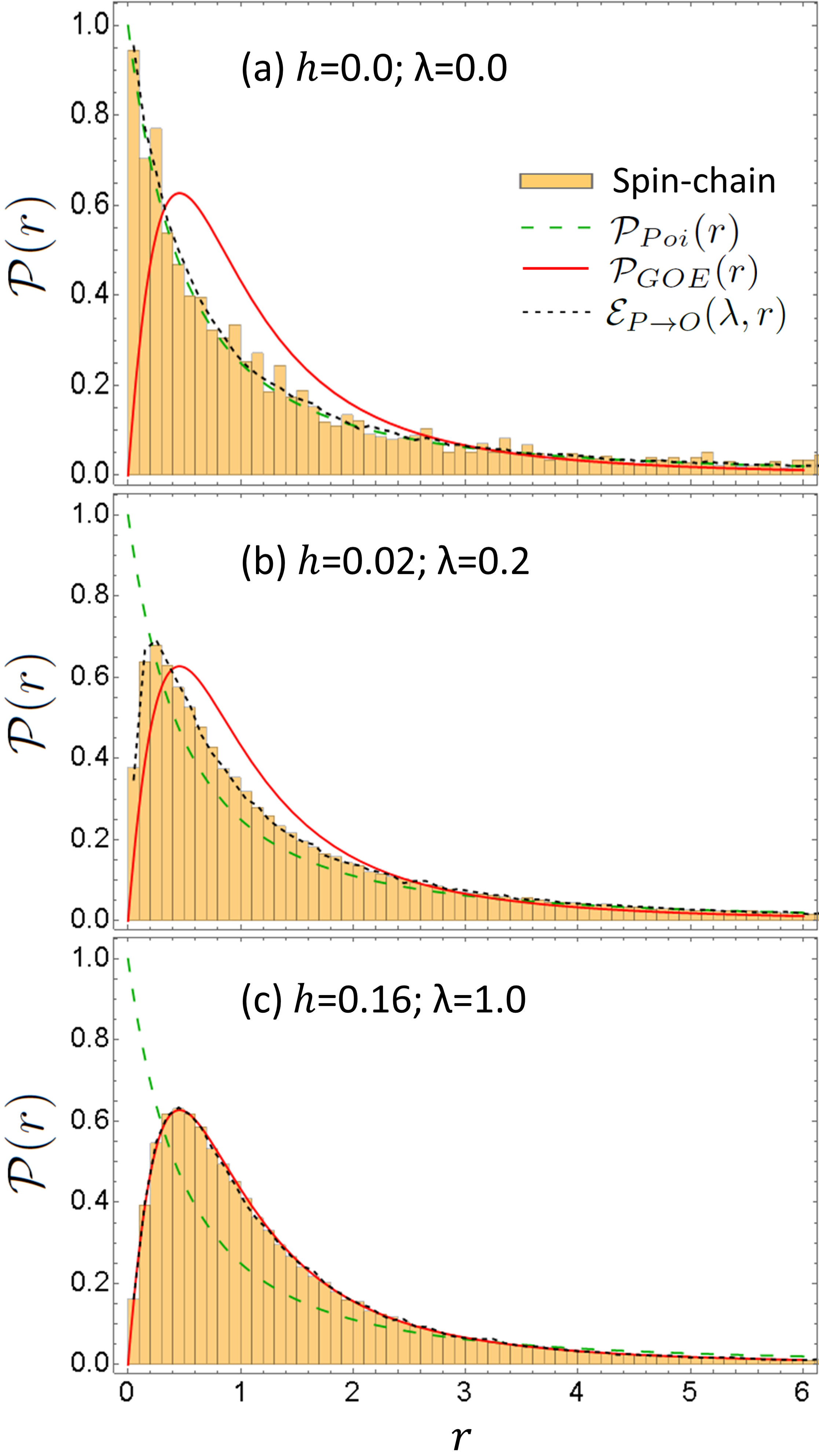}

\caption{(a) Poissonian $\rightarrow$ (b) Intermediate $\rightarrow$ (c) GOE crossover for probability distribution
of ratios with the increasing value of $\mathord{h}$ for the $\mathord{N=14}$
system.\label{fig:Poissonian-to-GOE_ratio_N=00003D14} }
\end{figure}

\begin{figure}[H]
\includegraphics[scale=0.2]{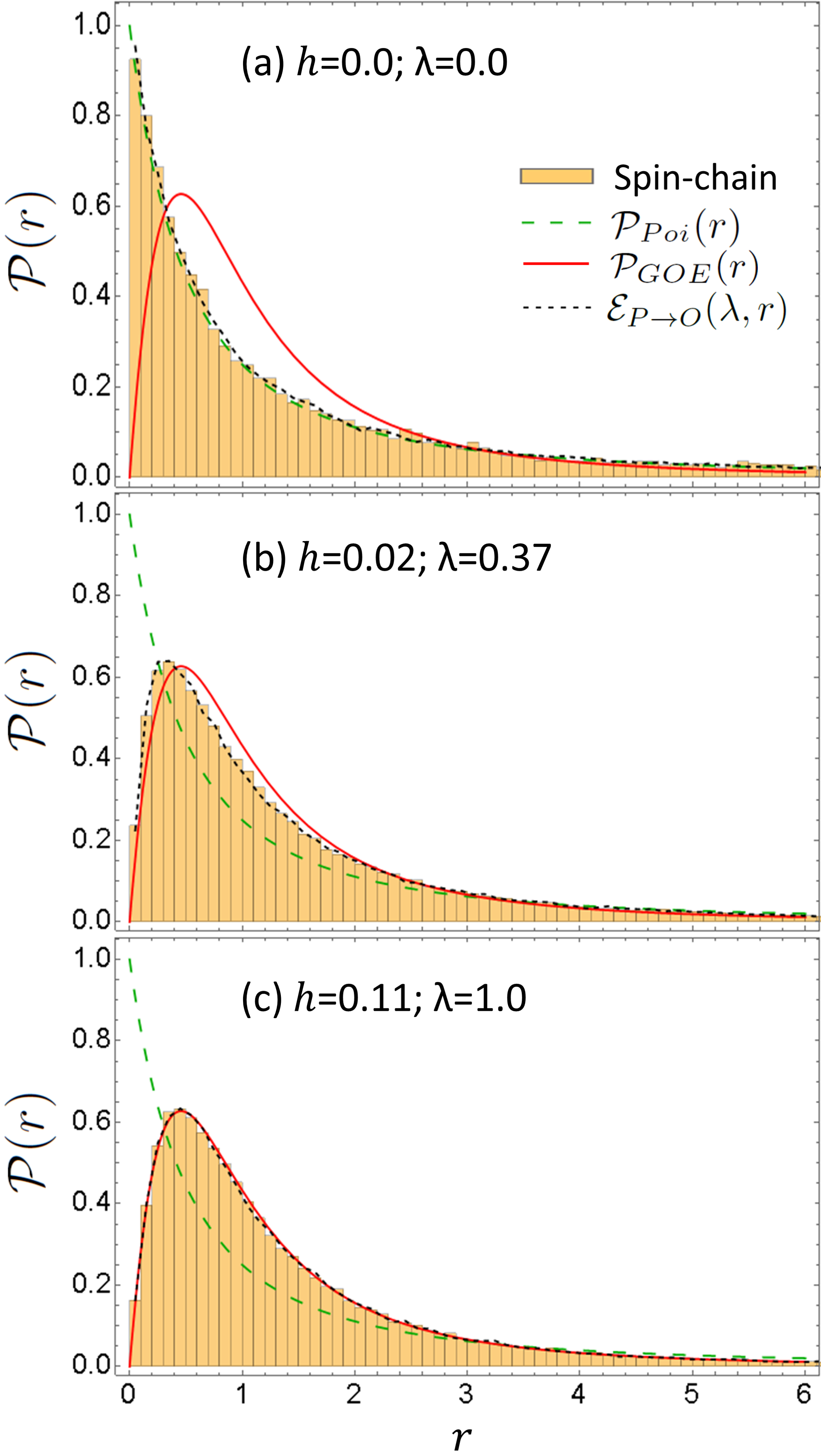}

\caption{(a) Poissonian $\rightarrow$ (b) Intermediate $\rightarrow$ (c) GOE crossover for probability distribution
of ratios with the increasing value of $\mathord{h}$ for the $\mathord{N=16}$
system.\label{fig:Poissonian-to-GOE_ratio_N=00003D16}}
\end{figure}

\begin{figure}
\includegraphics[scale=0.2]{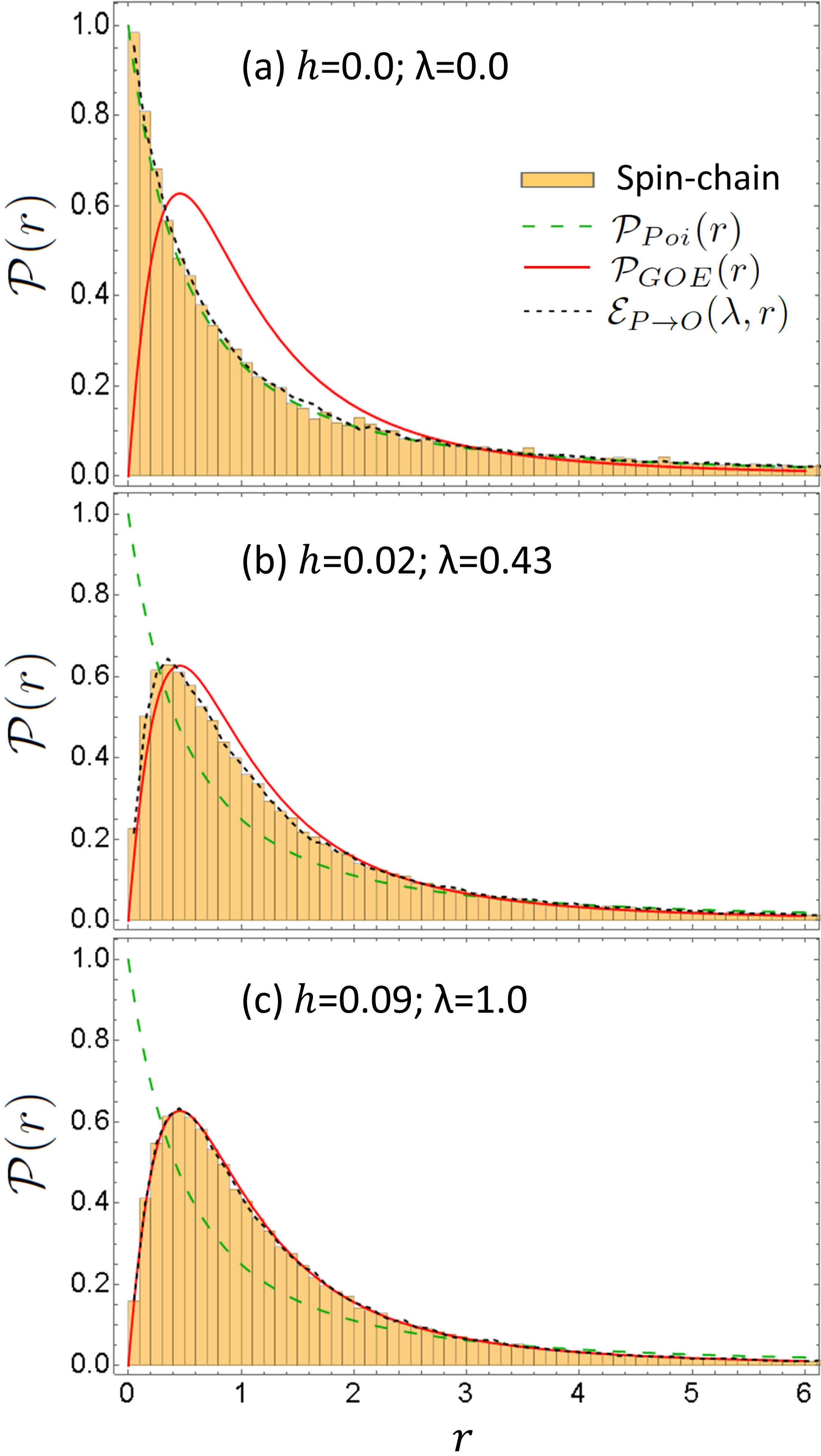}

\caption{(a) Poissonian $\rightarrow$ (b) Intermediate $\rightarrow$ (c) GOE crossover for probability distribution
of ratios with the increasing value of $\mathord{h}$ for the $\mathord{N=18}$
system. \label{fig:Poissonian-to-GOE_ratio_N=00003D18}}
\end{figure}

\subsubsection{Poissonian to GOE, and GOE to GUE Crossovers in Ratio Distribution \label{subsec:Both-crossovers-in-ratio-distribution}}

To study the Poissonian ($\mathcal{P}_{Poi}(r)$) to GOE ($\mathcal{P}_{GOE}(r)$)
crossover in the ratio distribution for our spin-chain
model (Eq. (\ref{eq:totalhamil})), we plot $\mathcal{P}(r)$ vs.
$\mathord{r}$ in Figs. \ref{fig:Poissonian-to-GOE_ratio_N=00003D14},
\ref{fig:Poissonian-to-GOE_ratio_N=00003D16} and \ref{fig:Poissonian-to-GOE_ratio_N=00003D18}
for the $\mathord{N=14}$ and $\mathord{N=16}$ and $N=18$ systems respectively.
We observe that for the $\mathord{N=14}$ system, the $\mathcal{P}(r)$ histogram
follows $\mathcal{P}_{GOE}(r)$ at $h=0.16$ whereas, for the $\mathord{N=16}$ and the $\mathord{N=18}$
systems the computed ratio distribution matches $\mathcal{P}_{GOE}(r)$ at around $h=0.11$ and $h=0.09$, respectively. These
values for the crossover points observed from NNSD, are quite comparable
with those obtained from the ratio distribution results. 

\begin{figure}
\includegraphics[scale=0.2]{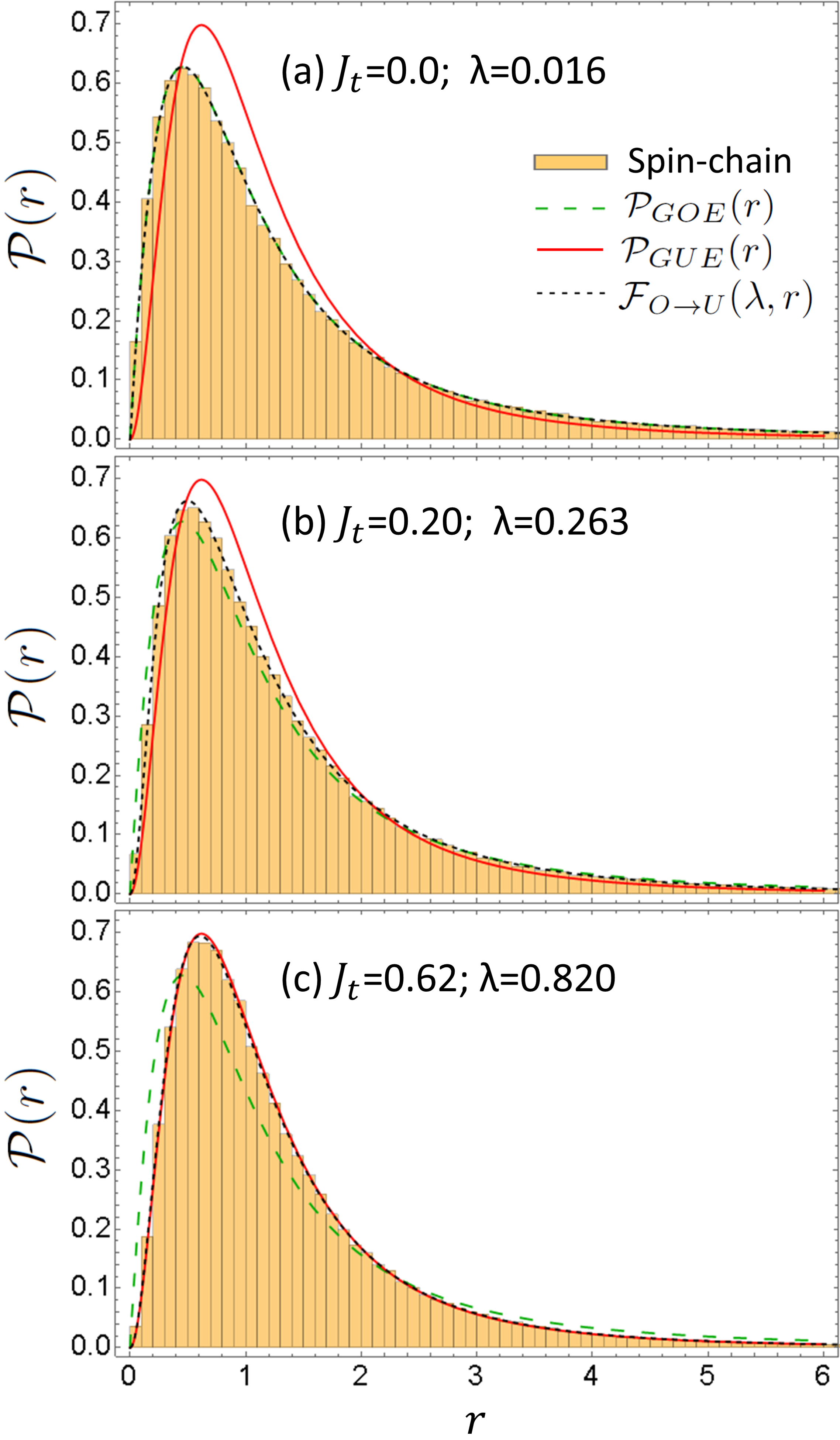}

\caption{\label{fig:GOE_to_GUE_ratio_N=00003D14}Ratio distribution for $N=14$;
(a), (b) and (c) shows the GOE to GUE crossover with the increasing
value of physical crossover parameter $J_{t}$, while the value
of the range of the random magnetic field ($h$) is kept at 0.2.
$\mathcal{F}_{\mathrm{\mathit{O}}\rightarrow U}(\lambda,r)$ interpolates
this crossover properly.}
\end{figure}

\begin{figure}
\includegraphics[scale=0.2]{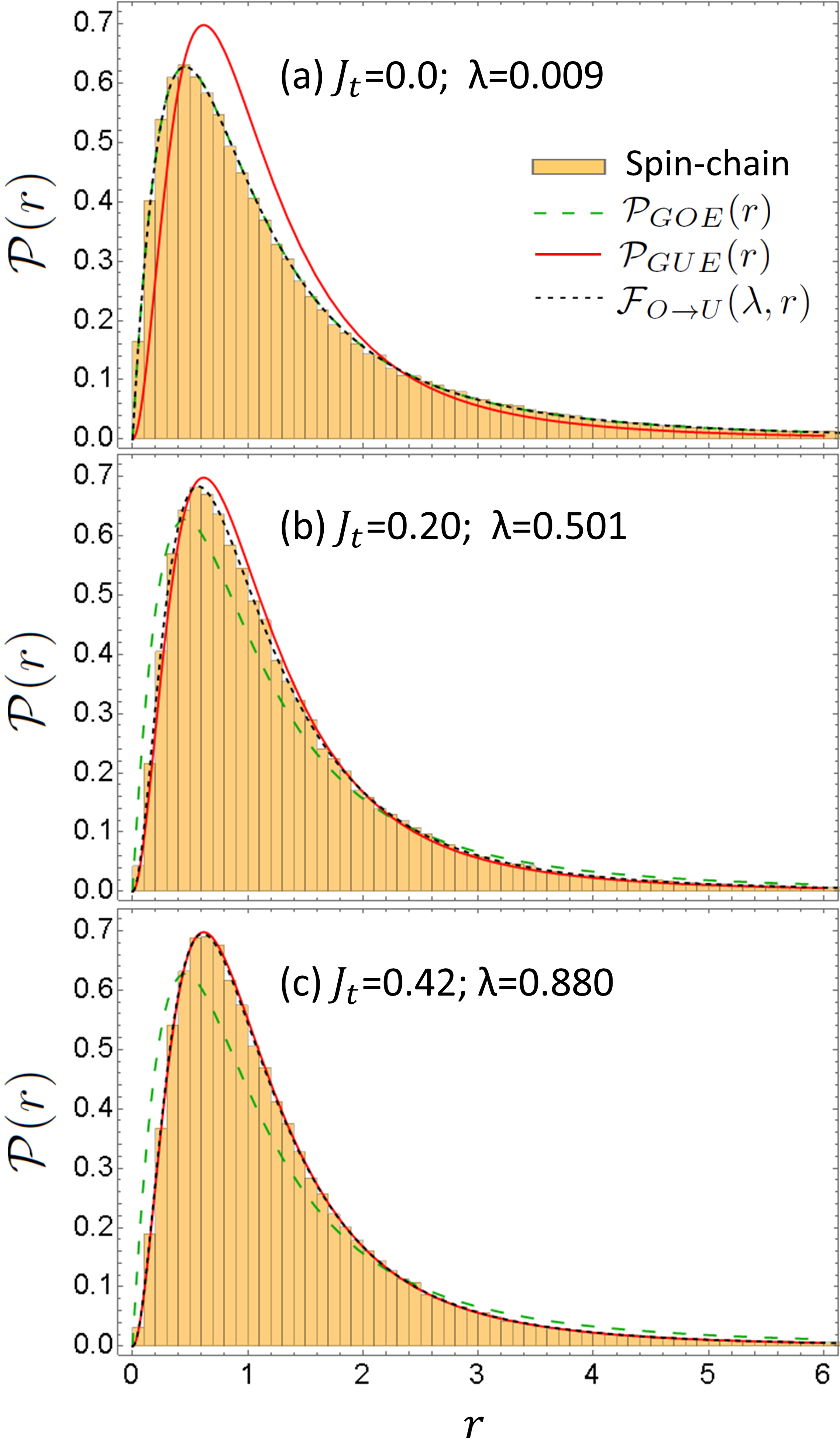}

\caption{\label{fig:GOE_to_GUE_ratio_N=00003D16}Ratio
distribution for the $N=16$ system, similar to the $N=14$ cases in Fig. \ref{fig:GOE_to_GUE_ratio_N=00003D14}.}
\end{figure}

\begin{figure}
\includegraphics[scale=0.2]{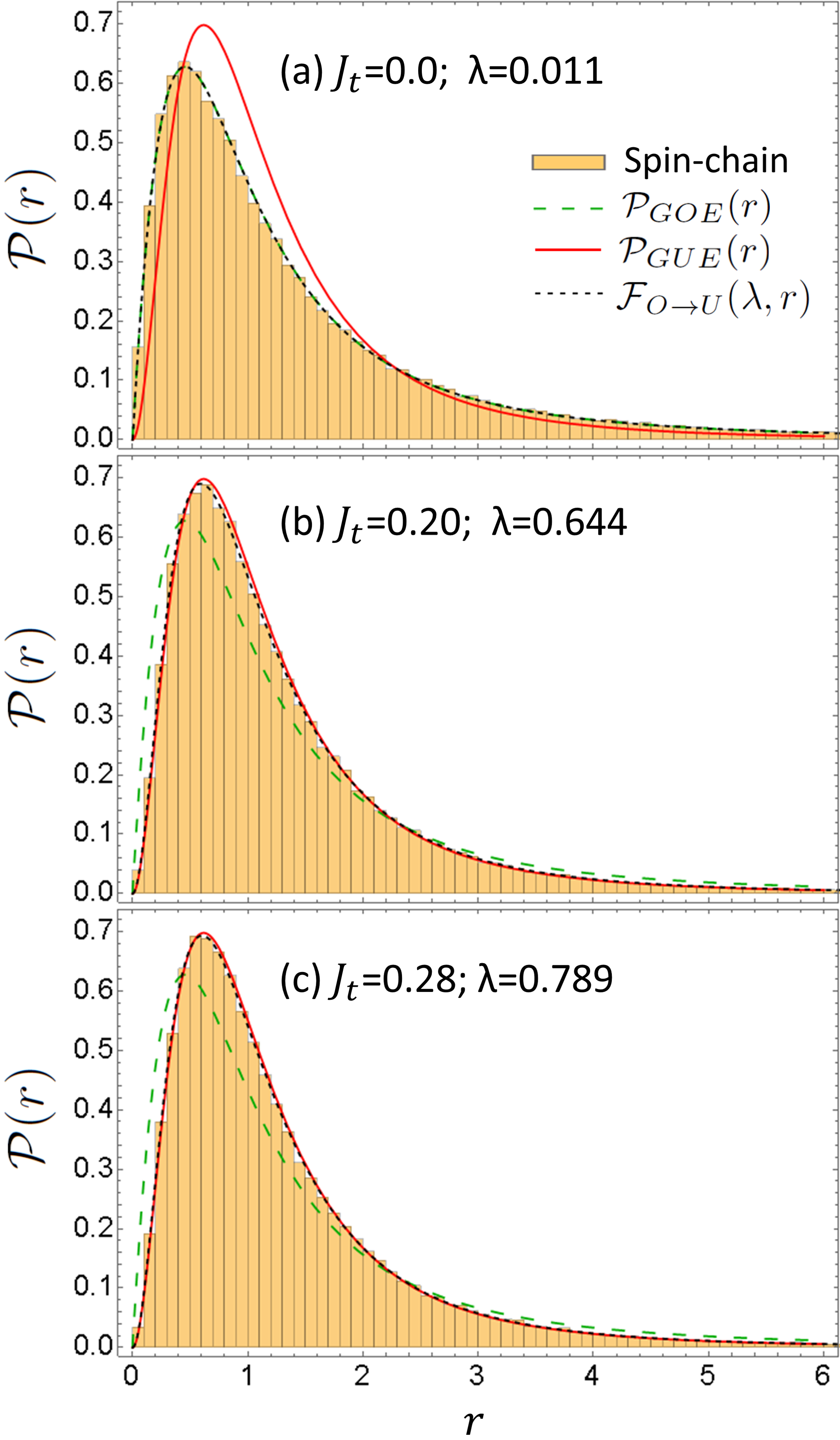}

\caption{\label{fig:GOE_to_GUE_ratio_N=00003D18}Ratio
distribution for the $N=18$ system, similar to the $N=14$ cases in Fig. \ref{fig:GOE_to_GUE_ratio_N=00003D14}}

\end{figure}

By keeping the spin-chain model well inside the GOE regime ($h=0.20$)
we study the GOE ($\mathcal{P}_{GOE}(r)$) to GUE ($\mathcal{P}_{GUE}(r)$)
crossover in the ratio distribution for the $N=14$ (Fig. \ref{fig:GOE_to_GUE_ratio_N=00003D14}),
$N=16$ (Fig. \ref{fig:GOE_to_GUE_ratio_N=00003D16}) and $N=18$ (Fig. \ref{fig:GOE_to_GUE_ratio_N=00003D18}) systems. The observed values
of scalar chirality parameter, $J_{t}$, at the GUE crossover in ratio distribution
match quite well with the NNSD findings, being at $J_{t}=0.62, 0.42$ and $0.28$ for the $N=14, 16$ and $18$ systems, as seen from Figs. \ref{fig:GOE_to_GUE_ratio_N=00003D14}, \ref{fig:GOE_to_GUE_ratio_N=00003D16} and \ref{fig:GOE_to_GUE_ratio_N=00003D18}, respectively.

For the Poissonian to GOE crossover, there is no analytical expression available for the ratio distribution interpolating function which can be compared with our observed results from spin-chains. Therefore, we simulate the Poissonian to GOE crossover using the Pandey-Mehta crossover Hamiltonian, as described in Eq. (\ref{eq:crossover_random_matrix}), using $3\times 3$ matrices $\mathcal{H_{\mathrm{0}}}$ and $\mathcal{H}_{\infty}$ taken from the Poissonian and the Gaussian Orthogonal ensembles, respectively. The crossover is realized 
by varying $\lambda$ from $0$ to relatively large numerical values. We calculate the probability distribution of ratios of the eigenvalue spacings for the matrix $\mathcal{H}$ and denote it by $\mathcal{E}_{P\rightarrow O}(\lambda,r)$.
We show in Figs. \ref{fig:Poissonian-to-GOE_ratio_N=00003D14}, \ref{fig:Poissonian-to-GOE_ratio_N=00003D16}
and \ref{fig:Poissonian-to-GOE_ratio_N=00003D18} that, $\mathcal{E}_{P\rightarrow O}(\lambda,r)$
can interpolate between the Poissonian to GOE crossover quite accurately for
our systems. 

For the GOE to GUE crossover, analytical expression for the interpolating function $\mathcal{F}_{\mathrm{\mathit{O}}\rightarrow U}(\lambda,r)$
(Eq. (\ref{eq:interpolate_ratio_GOE-GUE})) is available. We use it to map the GOE to GUE crossover
in our spin-chain model onto a RMT description. This function interpolates the crossover
perfectly for all the systems (Figs. \ref{fig:GOE_to_GUE_ratio_N=00003D14},
\ref{fig:GOE_to_GUE_ratio_N=00003D16} and \ref{fig:GOE_to_GUE_ratio_N=00003D18}).

The simulated interpolating probability distribution $\mathcal{E}_{P\rightarrow O}(\lambda,r)$
and the analytical interpolating function $\mathcal{F}_{\mathrm{\mathit{O}}\rightarrow U}(\lambda,r)$
are based on $3\times3$ matrix ensembles. But by applying these to a physical system here, we have successfully shown that, similar to Wigner surmises, they can
be used for much higher dimensional matrices, relevant for realistic physical systems, too.

\begin{figure}
\includegraphics[scale=0.2]{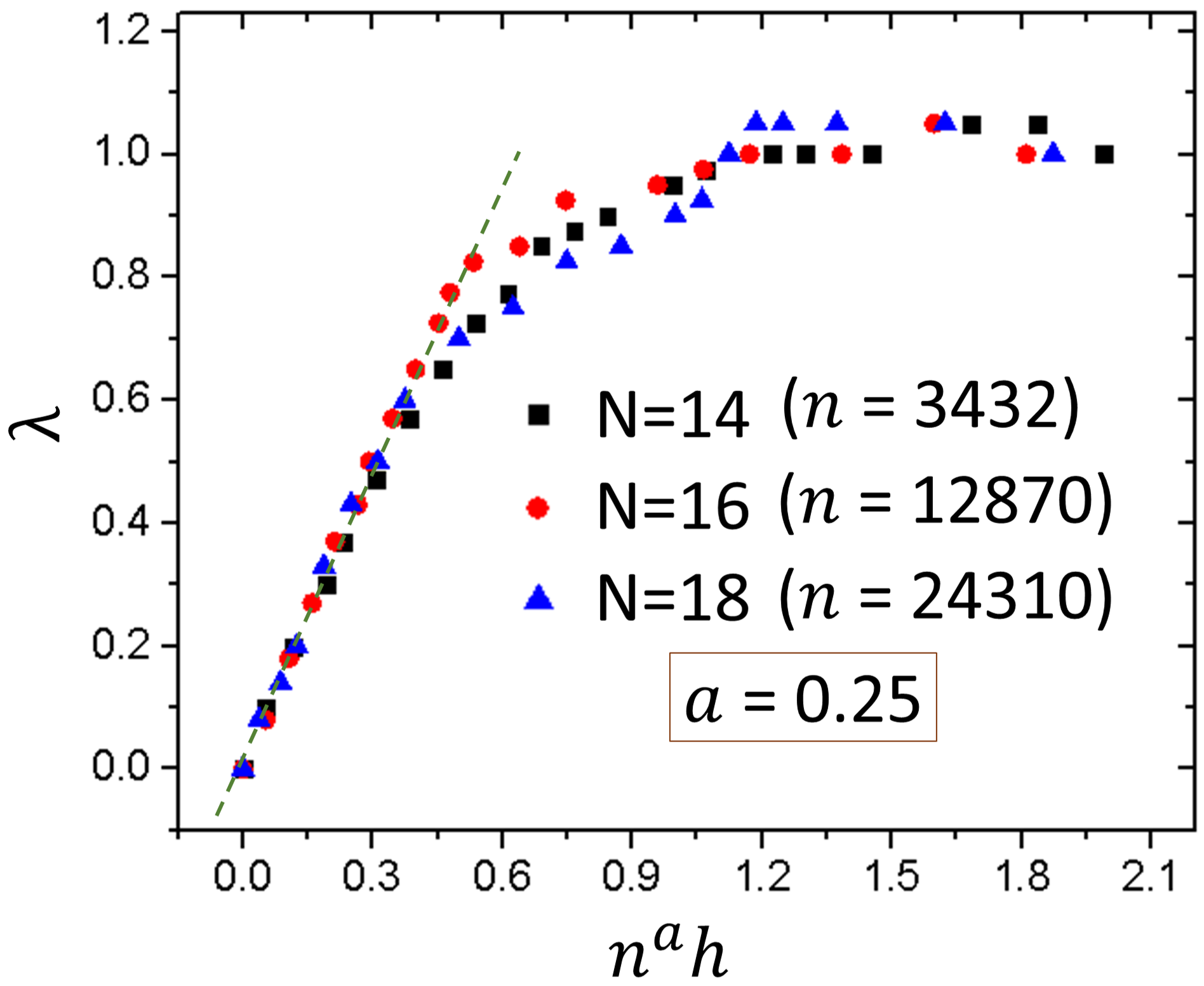}

\caption{\label{fig:Scaling-of-ratio-distribution-Poi-to-GOE}Scaling in ratio
distribution for the Poissonian to GOE crossover. $\lambda$ of the interpolating
matrix model's level-spacing ratio distribution, $\mathcal{E}_{P\rightarrow O}(\lambda,r)$,
is plotted against the scaled physical crossover parameter $h$. Here
the universal scaling exponent, $a=0.25.$ The dashed green line is
a fit based on the data points occurring in the linear regime. }

\end{figure}

\begin{figure}
\includegraphics[scale=0.2]{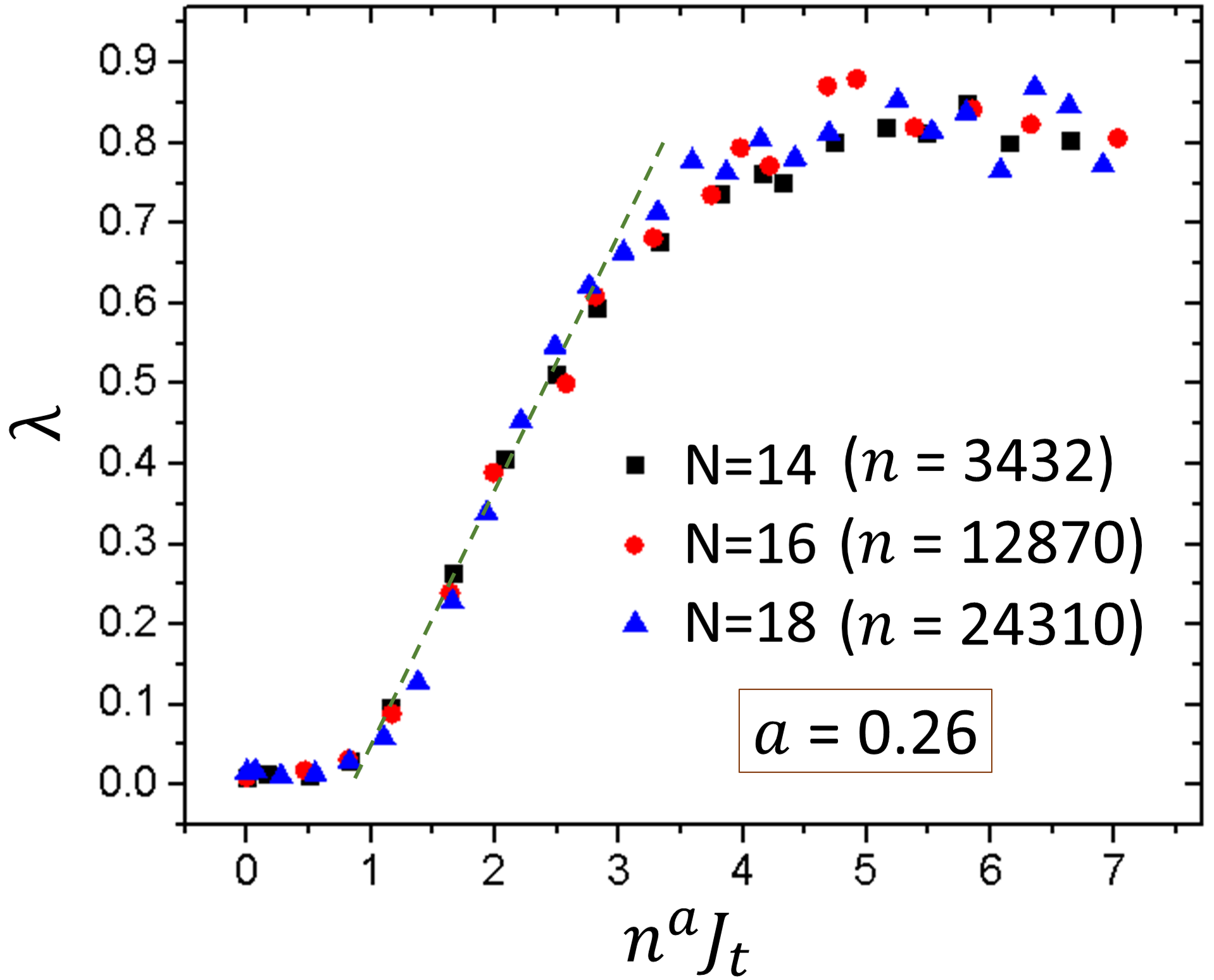}

\caption{\label{fig:Scaling-of-ratio-distribution-GOE-to-GUE}Scaling in ratio
distribution for the GOE to GUE crossover. $\lambda$ of the interpolating
function, $\mathcal{F}_{\mathrm{\mathit{O}}\rightarrow U}(\lambda,r)$, is
plotted against the scaled physical crossover parameter $J_{t}$.
Here the universal scaling exponent $a=0.26.$ The dashed green line
is a fit based on the data points occurring in the linear regime. }

\end{figure}

\subsubsection{Universality in Ratio Distribution \label{subsec:Universality in Ratio Distribution}}

In this section we study the universality aspects of both the crossovers in the ratio distribution. In the last section we studied the crossovers
in the ratio distribution $\mathcal{P}(r)$ with varying
physical crossover parameters $h$ and $J_{t}$. Here we have explored the variation of the generic RMT crossover parameter $\lambda$, against the physical
crossover parameters in the ratio distribution. The Poissonian to GOE
crossover in our system is interpolated quite accurately by the $\mathcal{E}_{P\rightarrow O}(\lambda,r)$
distribution, with increasing $\lambda$. In Fig. \ref{fig:Scaling-of-ratio-distribution-Poi-to-GOE}
we plot the variation of $\lambda$ with the physical crossover
parameter, $h$, scaled with $n^a$. We observe the emergence of universality in the Poissonian
to GOE crossover in the ratio distribution, for large system sizes. The
universal scaling exponent $a$ here assumes the value of $0.25$, which
is quite comparable to the corresponding universal scaling exponent of NNSD, where it has a value of 0.23. 

Since the ratio distribution for the GOE to GUE crossover in our system is properly
interpolated by the RMT interpolating function $\mathcal{F}_{\mathrm{\mathit{O}}\rightarrow U}(\lambda,r)$
with increasing $\lambda$, so we study the variation of $\lambda$ with
the scaled physical crossover parameter, $n^aJ_{t}$, in Fig. \ref{fig:Scaling-of-ratio-distribution-GOE-to-GUE}.
The universality of the GOE to GUE crossover in the ratio distribution is
observed in Fig. \ref{fig:Scaling-of-ratio-distribution-GOE-to-GUE}
and the universal scaling exponent comes out to be $a=0.26.$
This is very close to the universal exponent of $0.25$, observed
in NNSD, pointing towards a universality in the value of the scaling exponent itself.

\section{Conclusions\label{sec:Conclusion}}

We have studied the eigenvalue statistics and spectral crossovers between different RMT symmetry classes (Poissonian $\rightarrow$ GOE and GOE $\rightarrow$ GUE), for 1D quantum spin chains with multi-spin interactions, coupled to random, inhomogeneous magnetic fields, as various symmetries are gradually broken by tuning of suitable physical parameters. We have focussed on its correspondence with the corresponding crossovers in exact RMT models, as embodied in the NNSD (which captures a two-point spectral correlation) and the RD (which reflects a three-point spectral correlation). We find an excellent one-to-one correspondence between the RMT results and the properties of the physical spin systems, which points towards the robustness of the RMT formalism and the Wigner-Dyson-Berry-Tabor classification, as also encountered in many other physical systems \cite{Avishai-Richert-PRB,Modak's-paper_2014,Wigner-surmise-for-high-orde-Abul-Magd,Chavda-Kota-PLA-1,Bertrand_spin_chain_ratio_1,Iyer_Oganesyan_quasi-periodic_system,Collura_Bose_Hubbard_2,Pandey1988,metallic_ring_in_mag_field,Pragya_Shukla_metal_insulator_transition,magneto-conductance_ballistic_quantum_dots,bipartite_entanglement_entropy}. We summarize below some salient conclusions from this detailed study :

(1) To obtain a good description in terms of RMT results, one needs relatively large lattice sizes, which luckily saturates soon enough by $N=18$. One also finds that the symmetry is broken by progressively smaller symmetry breaking physical parameters, as the system size is increased.

(2) Using a finite-size scaling analysis we have uncovered a Universality in these crossovers. There are two facets to this. {\em First}, by suitably scaling the physical crossover parameter by the Hamiltonian basis size raised to the power of a suitable scaling exponent ($a$), in each type of crossover, one is able to collapse the RMT scaling parameter ($\lambda$) {\em vs.} the physical scaling parameter plots for various lattice sizes, onto one universal plot. {\em Second}, irrespective to the type of crossover (Poissonian $\rightarrow$ GOE, or GOE $\rightarrow$ GUE) or the spectral correlation measure used (i.e. NNSD or RD), the scaling exponent ($a$) seems to hover around the value of $\sim0.25$ (between $0.23$ - $0.26$). Interestingly, the value of this exponent seems to be quite robust too.

(3) For this kind of investigation, the many-body basis size grows very fast with the lattice size, in spite of restricting to the smallest $S^{z}$ sector. Moreover, the exact diagonalization process needs to be repeated a large number of times, as physical parameters have to be continuously varied over a sizeable range to capture the crossovers and also for configuration (ensemble) averaging over the random magnetic field. To add to the problem, one typically needs {\em all} eigenvalues, and so one cannot rely on iterative diagonalization schemes like the Lanczos method to reduce computational costs as they yield only extremal eigenvalues and eigenstates. The only saving grace is the use of additional symmetries and we have demonstrated here the usefulness of one such symmetry, {\em viz.} global spin-rotational symmetry to truncate to an {\em inverted basis}, that reduces the basis size to half its original value.

Future directions could involve extending these methods to study higher-dimensional correlated spin models, long-range spectral fluctuations or generalization to Fermionic models etc. Some of these investigations are already under way. 

\begin{acknowledgments}
DK and SSG acknowledge financial support from the DST-SERB Project ECR/2016/002054.
\end{acknowledgments}

\appendix* \section{Simplification of the Scalar Spin Chirality Term}\label{sec:simplification-of-chirality-term​}

This appendix presents a compact derivation of Eq. (\ref{eq:simplified_scalar_chirality}).
Einstein summation convention is used throughout this section. We have,
\begin{align}
\label{eq:Scalar_chirality_appendix_1}
\nonumber
\mathbf{S}_{j}\cdot[\mathbf{S}_{j+1}\times\mathbf{S}_{j+2}]=\mathbf{S}_{j}\cdot\mathbf{S}_{c}=\mathbf{\mathrm{S}}_{j}^{\alpha}\mathbf{\mathrm{S}}_{c}^{\alpha}\\
=\varepsilon_{\alpha\beta\gamma}\mathbf{\mathrm{S}}_{j}^{\alpha}\mathbf{\mathrm{S}}_{(j+1)}^{\beta}\mathbf{\mathrm{S}}_{(j+2)}^{\gamma}.
\end{align}
Here, $\alpha$, $\beta$, and $\gamma$ run over the spin components $\mathrm{(x,y,z)}$. As we know exchanging rows and columns leaves a determinant unchanged, we obtain,
\begin{align}
\label{eq:Scalar_chirality_appendix_2}
\nonumber
\varepsilon_{\alpha\beta\gamma}\mathbf{\mathrm{S}}_{j}^{\alpha}\mathbf{\mathrm{S}}_{(j+1)}^{\beta}\mathbf{\mathrm{S}}_{(j+2)}^{\gamma}&=\left|\begin{array}{ccc}
\mathbf{\mathrm{S_{\mathit{j}}^{x}}} & \mathbf{\mathrm{S_{\mathit{j}}^{y}}} & \mathbf{\mathrm{S_{\mathit{j}}^{z}}}\\
\mathbf{\mathrm{S_{(\mathit{j}+1)}^{x}}} & \mathbf{\mathrm{S_{(\mathit{j}+1)}^{y}}} & \mathbf{\mathrm{S_{(\mathit{j}+1)}^{z}}}\\
\mathbf{\mathrm{S_{(\mathit{j}+2)}^{x}}} & \mathbf{\mathrm{S_{(\mathit{j}+2)}^{y}}} & \mathbf{\mathrm{S_{(\mathit{j}+2)}^{z}}}
\end{array}\right|\\
\nonumber
&=\left|\begin{array}{ccc}
\mathbf{\mathrm{S_{\mathit{j}}^{x}}} & \mathbf{\mathrm{S_{(\mathit{j}+1)}^{x}}} & \mathbf{\mathrm{S_{(\mathit{j}+2)}^{x}}}\\
\mathbf{\mathrm{S_{\mathit{j}}^{y}}} & \mathbf{\mathrm{S_{(\mathit{j}+1)}^{y}}} & \mathrm{S_{(\mathit{j}+2)}^{y}}\\
\mathbf{\mathrm{S_{\mathit{j}}^{z}}} & \mathbf{\mathrm{S_{(\mathit{j}+1)}^{z}}} & \mathbf{\mathrm{S_{(\mathit{j}+2)}^{z}}}
\end{array}\right|\\
&=\varepsilon_{klm}\mathrm{S}_{k}^{\mathrm{x}}\mathrm{S}_{l}^{\mathrm{y}}\mathrm{S}_{m}^{\mathrm{z}},
\end{align}
where $k=j,$ $l=j+1$ and $m=j+2$. With this the Levi-Civita indices now run over sites, rather than spin components $\mathrm{(x,y,z)}$.
 
The spin components are related to spin ladder operators as, 
\begin{align}
\label{eq:Scalar_chirality_appendix_3}
\mathbf{\mathrm{S^{x}}}=\frac{1}{2}(\mathbf{\mathrm{S}}^{+}+\mathbf{\mathrm{S}}^{-}),~~~
\mathbf{\mathrm{S^{y}}}=\frac{1}{2i}(\mathbf{\mathrm{S}}^{+}-\mathbf{\mathrm{S}}^{-}).
\end{align}
Now, using Eq. (\ref{eq:Scalar_chirality_appendix_3}) we get,
\begin{align}
\nonumber
\varepsilon_{klm}\mathrm{S}_{k}^{\mathrm{x}}\mathrm{S}_{l}^{\mathrm{y}}\mathrm{S}_{m}^{\mathrm{z}}&=\frac{1}{4i}[\varepsilon_{klm}\mathbf{\mathrm{S_{\mathit{k}}^{+}}}\mathbf{\mathrm{S_{\mathit{l}}^{+}}}\mathrm{S}_{m}^{\mathrm{z}}-\varepsilon_{klm}\mathbf{\mathrm{S_{\mathit{k}}^{-}}}\mathbf{\mathrm{S_{\mathit{l}}^{-}}}\mathrm{S}_{m}^{\mathrm{z}}\\
&-\varepsilon_{klm}\mathrm{S_{\mathit{k}}^{+}}\mathbf{\mathrm{S_{\mathit{l}}^{-}}}\mathrm{S}_{m}^{\mathrm{z}}+\varepsilon_{klm}\mathrm{S_{\mathit{k}}^{-}}\mathbf{\mathrm{S_{\mathit{l}}^{+}}}\mathrm{S}_{m}^{\mathrm{z}}].\label{eq:Scalar_chirality_appendix_4}
\end{align}

The first term leads to
\begin{align}
\nonumber
\varepsilon_{klm}\mathbf{\mathrm{S_{\mathit{k}}^{+}}}\mathbf{\mathrm{S_{\mathit{l}}^{+}}}\mathrm{S}_{m}^{\mathrm{z}}=-\varepsilon_{lkm}\mathbf{\mathrm{S_{\mathit{k}}^{+}}}\mathbf{\mathrm{S_{\mathit{l}}^{+}}}\mathrm{S}_{m}^{\mathrm{z}}\\
=-\varepsilon_{lkm}\mathbf{\mathrm{S_{\mathit{l}}^{+}}}\mathbf{\mathrm{S_{\mathit{k}}^{+}}}\mathrm{S}_{m}^{\mathrm{z}}=-\varepsilon_{klm}\mathbf{\mathrm{S_{\mathit{k}}^{+}}}\mathbf{\mathrm{S_{\mathit{l}}^{+}}}\mathrm{S}_{m}^{\mathrm{z}},
\end{align}
 where in the second step we have used the commutation relation $\left[\mathbf{\mathrm{S_{\mathit{k}}^{+}}},\mathbf{\mathrm{S_{\mathit{l}}^{+}}}\right]=0$
between distinct sites, and in the last step we have exchanged the
dummy indices $k$ and $l$. Thus, it follows that
\[
\varepsilon_{klm}\mathbf{\mathrm{S_{\mathit{k}}^{+}}}\mathbf{\mathrm{S_{\mathit{l}}^{+}}}\mathrm{S}_{m}^{\mathrm{z}}=0.
\]
Similarly, we have
\[
\varepsilon_{klm}\mathbf{\mathrm{S_{\mathit{k}}^{-}}}\mathbf{\mathrm{S_{\mathit{l}}^{-}}}\mathrm{S}_{m}^{\mathrm{z}}=0.
\]
Now, the third term of Eq. (\ref{eq:Scalar_chirality_appendix_4}) leads to
\begin{align*}
-\varepsilon_{klm}\mathbf{\mathrm{S_{\mathit{k}}^{+}}}\mathbf{\mathrm{S_{\mathit{l}}^{-}}}\mathrm{S}_{m}^{\mathrm{z}}=\varepsilon_{lkm}\mathbf{\mathrm{S_{\mathit{k}}^{+}}}\mathbf{\mathrm{S_{\mathit{l}}^{-}}}\mathrm{S}_{m}^{\mathrm{z}}\\
=\varepsilon_{lkm}\mathbf{\mathrm{S_{\mathit{l}}^{-}}}\mathbf{\mathrm{S_{\mathit{k}}^{+}}}\mathrm{S}_{m}^{\mathrm{z}}=\varepsilon_{klm}\mathbf{\mathrm{S_{\mathit{k}}^{-}}}\mathbf{\mathrm{S_{\mathit{l}}^{+}}}\mathrm{S}_{m}^{\mathrm{z}},
\end{align*}
 where in the second step we have used the commutation relation $\left[\mathbf{\mathrm{S_{\mathit{k}}^{+}}},\mathbf{\mathrm{S_{\mathit{l}}^{-}}}\right]=0$, because the Levi-Civita symbol ensures that $k$ and $l$ always assume distinct site values,
and in the last step we have exchanged the dummy indices $k$ and
$l$. So, Eq. (\ref{eq:Scalar_chirality_appendix_4}) becomes,
\begin{align*}
\varepsilon_{klm}\mathrm{S}_{k}^{\mathrm{x}}\mathrm{S}_{l}^{\mathrm{y}}\mathrm{S}_{m}^{\mathrm{z}}=\frac{1}{4i}2\varepsilon_{klm}\mathbf{\mathrm{S_{\mathit{k}}^{-}}}\mathbf{\mathrm{S_{\mathit{l}}^{+}}}\mathrm{S}_{m}^{\mathrm{z}}\\
=\frac{-i}{2}\varepsilon_{klm}\mathbf{\mathrm{S_{\mathit{k}}^{-}}}\mathbf{\mathrm{S_{\mathit{l}}^{+}}}\mathrm{S}_{m}^{\mathrm{z}}=\frac{i}{2}\varepsilon_{mlk}\mathrm{S}_{m}^{\mathrm{z}}\mathbf{\mathrm{S_{\mathit{k}}^{-}}}\mathbf{\mathrm{S_{\mathit{l}}^{+}}}\\
=\frac{i}{2}\varepsilon_{mlk}\mathrm{S}_{m}^{\mathrm{z}}\mathbf{\mathrm{S_{\mathit{l}}^{+}}}\mathbf{\mathrm{S_{\mathit{k}}^{-}}}=\frac{i}{2}\varepsilon_{klm}\mathrm{\mathrm{S}_{\mathit{k}}^{\mathrm{z}}\mathrm{\mathbf{\mathrm{S_{\mathit{l}}^{+}}}\mathbf{\mathrm{S_{\mathit{m}}^{-}}}},}
\end{align*}
where the names of dummy indices have just been interchanged. This
establishes Eq. (\ref{eq:simplified_scalar_chirality}).

\bibliographystyle{apsrev4-2}

\end{document}